\documentclass{aa}

\usepackage{graphicx}
\usepackage{txfonts}
\usepackage{bm}
\usepackage[breaklinks=true]{hyperref}
\usepackage{xcolor}
\hypersetup{
    colorlinks,
    linkcolor={blue!50!black},
    citecolor={blue!50!black},
    urlcolor={blue!50!black}
}
\usepackage{natbib}
\bibpunct{(}{)}{;}{a}{}{,}
\usepackage{lscape}

\newcommand{\Msun}{\rm M_{\sun}}

\newcommand{\kpc}{\rm kpc}

\newcommand{\cMpc}{\rm cMpc}

\newcommand{\lsim}{\mathrel{\hbox{\rlap{\lower.55ex\hbox{$\sim$}} \kern-.3em\raise.4ex\hbox{$<$}}}}
\newcommand{\gsim}{\mathrel{\hbox{\rlap{\lower.55ex\hbox{$\sim$}} \kern-.3em\raise.4ex\hbox{$>$}}}}

\newcommand{\Ato}{A_{2} }

\newcommand{\Amax}{A_{\rm 2,max} }

\newcommand{\rbar}{r_{\rm bar} }
\newcommand{\rhalf}{r_{\rm 50,*} }

\newcommand{\Dthin}{(D/T)_{\rm thin}}

\newcommand{\Dthinzo}{(D/T)_{\rm thin,z0}}
\newcommand{\Dthinzu}{(D/T)_{\rm thin,z1}}

\newcommand{\tbar}{t_{\rm bar} }

\newcommand{\TNGF}{TNG50}

\defcitealias{rosasguevara2020}{RG20}
\defcitealias{rosasguevara2022}{RG22}

\begin{document}

   \title{The rise and fall of bars in disc galaxies from $z=1$ to $z=0$}

   \subtitle{The role of environment}

     \author{Yetli Rosas-Guevara\inst{1}\thanks{email:yetli.rosas@dipc.org}
          \and
          Silvia Bonoli \inst{1}\fnmsep\inst{2}
          \and
          Carmen Misa Moreira\inst{3}
          \and
          David Izquierdo-Villalba\inst{4} \fnmsep\inst{5}
          }
   \institute{Donostia International Physics Centre (DIPC), Paseo Manuel de Lardizabal 4, 20018 Donostia-San Sebastian, Spain
         \and
             IKERBASQUE, Basque Foundation for Science, E-48013, Bilbao, Spain
         \and
             Valencian International University (VIU), Calle Pintor Sorolla 21,
46002 Valencia, Spain
         \and
             Dipartimento di Fisica G. Occhialini, Universit\`{a} di Milano-Bicocca, Piazza della Scienza 3, IT-20126 Milano, Italy
         \and
             INFN, Sezione di Milano-Bicocca, Piazza della Scienza 3, IT-20126 Milano, Italy
             }

   \date{Received September 15, 1996; accepted March 16, 1997}


  \abstract
  {Stellar bars are non-axisymmetric structures found in over $30$ per cent of massive disc galaxies in the local Universe.  The environment could play a significant role in determining whether or not a spiral galaxy is likely to develop a bar.}
   {We investigate the influence of the environment on the evolution of barred and unbarred disc galaxies with a mass of larger than $10^{10}\Msun$ from $z=1$ down to $z=0$, employing the TNG50 magnetic-hydrodynamical simulation.}
   {We determined the fraction of barred galaxies that conserve their bar and the fraction of those that lost it by $z=0$.
   We also estimate the fraction of unbarred galaxies at z=1 that develop a bar at later times. We study the merger histories and the distance of close companions for each category to understand the role of the environment in the evolution of these galaxies. }
   {We find that 49 per cent of $z=1$ disc galaxies undergo a morphological transformation, transitioning into either a lenticular or spheroidal galaxy, while the other $51$ per cent retain the large disc shape. The morphological alteration is mostly influenced by the environment. Lenticular and spheroidal galaxies tend to exist in denser environments and have more frequent mergers compared to disc galaxies. We find that bars are stable after they have formed, as over half of the barred galaxies (60.2 per cent) retain the bar structure and have experienced fewer mergers compared to those galaxies that lose their bars (5.6 per cent). These latter galaxies start with weaker and shorter bars at $z=1,$  are influenced by tidal interactions, and are frequently observed in more populated areas. Additionally, our study reveals that less than 20 per cent of unbarred galaxies will never develop a bar and exhibit the quietest merger history. Unbarred galaxies that undergo bar formation after $z=1$   more frequently experience a merger event.  Furthermore, tidal interactions with a close companion may account for bar formation in at least one-third of these instances.}
   {Our findings highlight that stable bars are prevalent in disc galaxies. Bar evolution may nonetheless be affected by the environment. Interactions with nearby companions or tidal forces caused by mergers have the capacity to disrupt the disc. This perturbance may materialise as the dissolution of the bar, the formation of a bar, or, in its most severe form, the complete destruction of the disc, resulting in morphological transformation. Bars that are weak and short at $z=1$ and undergo major or minor mergers may eventually dissolve, whereas unbarred galaxies that enter crowded environments or experience a merger may develop a bar. }



   \maketitle
%

\section{Introduction}

Stellar bars are non-axisymmetric structures that are found in the central regions of the majority of disc galaxies. They are present in over $30$ per cent of massive disc galaxies with a stellar mass of $10^{10} M_{\odot}$  in the local Universe (e.g. \citealt{sellwood1993,masters2011,gavazzi2015}). These bars are believed to be drivers of the secular evolution of galaxies, as they are efficient in redistributing gas, stars, and dark matter towards the central regions of the galaxy (e.g. \citealt{debattista2004, athanassoula2005}). 

The presence or absence of bars in galaxies can reveal crucial information about the assembly history of these galaxies. One effective way to investigate this is by studying the evolution of bar fractions as a function of various galaxy properties, such as stellar mass, gas content, or disc structure. Local Universe studies have shown that bars are more frequently found in massive, gas-poor, and red galaxies than in their blue, gas-rich, and star-forming counterparts  (e.g. \citealt{barazza2008,masters2012,gavazzi2015,consolandi2016,cervantes2017}). Moreover, the redshift dependence of the bar fraction can provide additional insights into the evolution of bars and their host galaxies.

Previous studies determined the bar frequency up to redshift $z=1$. Earlier studies, based on small samples of disc galaxies, found contradictory trends in the evolution of the bar fraction. \cite{abraham1999} found a decreasing bar fraction with increasing redshift using Hubble Space Telescope (HST) observations, while \cite{elmegreen2004}, and \cite{jogee2005} suggested a constant fraction. Later, \cite{sheth2008}, using a larger sample of discs from the Cosmic Evolution Survey (COSMOS), found that the fraction of barred disc galaxies rapidly declines with increasing redshift, from $\sim 0.65$ at $z\sim 0.2$ to $\sim 0.20$ at $z\sim 0.8$. \cite{melvin2014} observed a similar decreasing trend in the bar fraction using a different selection of disc galaxies from COSMOS and with a different bar-identification method provided by the Galaxy Zoo Hubble (GZH) project.
Notably,  \cite{erwin2018}, who studied a sample of spiral galaxies from the Spitzer Survey of Stellar Structure in Galaxies (S$^4$G), suggested that bar fractions at high redshift may be underestimated and \cite{menendez2023} pointed out that the differences in the bar properties when observed in different bands could also affect the results of high-redshift studies of bars. Recently, with the arrival of observational facilities such as JWST, it has become possible to detect barred galaxies to even higher redshifts \citep{guo2023,tsukui2023,leconte2023,costantin2023}, providing new constraints on the evolution and formation of bar structures.


Explorations of whether or not the bar fraction is increasing or decreasing can be used to study the lifetimes of bar structures providing useful information about whether these structures are stable in time or disappear through secular processes \cite*[e.g.][]{aguerri2009} or through external processes \cite*[e.g.][]{zana2018b, peschken2019,mendez2023}.  This topic is hotly debated.  The secular process of forming and dissolving a stellar bar can be complex and takes place over long timescales. Bars can be formed spontaneously via secular evolution as a result of global, non-axisymmetric instabilities \cite*[e.g.][]{athanassoula2003,mendez2012,gavazzi2015}.  On the other hand, the combined effects of central mass concentrations (CMCs) and gravity torques could make galactic bars transient features, with lifetimes of 1-2 Gyr in typical Sb-Sc galaxies  \citep{bournaud2005}.

The formation of new stars can lead to feedback processes that release energy and momentum that can disrupt the bar structure and gradually weaken and dissolve it \cite*[e.g.][]{zana2018c}. Vertical resonances can  also arise within the bar, causing stars to oscillate vertically within the bar. This can lead to the formation of a central mass concentration, which can destabilise the bar and cause it to dissolve \cite*[e.g.][]{martinez2006}.

In the case of external processes, galactic mergers and interactions can play a significant role in the formation and disruption of stellar bars in galaxies \cite[e.g.][]{elmegreen1990,berentzen2004,lokas2016,zana2018a,zana2018b,lokas2019}.  The specific conditions required for a galactic merger or interaction to form or dissolve a bar include the mass ratio of the merging galaxies, and the phase of the bar and of the orbit, because the strength of the perturbation can be a deciding factor in whether or not the central part of the galaxy can form a bar \citep{elmegreen1990}. The orbit of the merging galaxies can be another condition; a direct collision between two galaxies can cause significant perturbations that can lead to the formation or dissolution of a bar \citep{berentzen2004}. Moreover, prograde encounters (when the angular momenta of the two galaxies are aligned) have a much more dramatic effect on the galactic structure than retrograde ones (when the angular momenta point in opposite directions). Prograde encounters lead to the formation of long and narrow tidal arms and tidally induced bars in galaxies orbiting a larger host, but retrograde ones do not \citep{lokas2018}. The gas content of the merging galaxies also contributes to the dissolution of a bar. If the merging galaxies have a significant amount of gas for instance, tidal interactions due to minor mergers can weaken the central stellar bar in the merger remnant  \citep{ghosh2021}.



Tidal forces due to dense environments or close companions also have a significant impact on the formation or dissolution of bars. \cite{lokas2014} and \cite{lokas2016} study the formation of a bar in galaxy encounters with a Milky-way galaxy and dwarf galaxy embedded in dense environments using N-body simulations. The authors find that the characteristics of the bars undergo temporal variations and are subject to the influence of the magnitude of the tidal force encountered. The formation of bars in galaxies is observed to occur at earlier stages and the bar itself is of greater strength and length in galaxies that experience a higher degree of tidal force exerted by the surrounding cluster.

It has very recently become possible to study the external and internal processes of galaxies in detail  with zoomed-in simulations (e.g. \citealt{kraljic2012,scannapieco2012,bonoli2016}).
In particular, \cite{zana2018a,zana2018b}, studying the \textit{Eris} simulation, focus on the effects of unequal-mass flybys and interactions. These authors find that flybys and mergers have the potential to impact the formation of a bar by inducing a delay in the time of its formation but do not bring about any significant alterations to the overall bar characteristics.

The study of the evolution of barred galaxies and their properties in a statistical framework is now also possible thanks to cosmological hydrodynamic simulations (\citealt{vogelsberger2014a,schaye2015,pillepich2018b,nelson2018}). Analysing the EAGLE simulation, \cite{algorry2017} discovered that bars slow down very quickly as they evolve, expanding the inner parts of the dark matter halo.  \cite{rosasguevara2020} studied massive barred disc galaxies at $z=0$ in the TNG100 simulation (see also \citealt{peschken2019,zhao2020,zhou2020} for Illustris and IllustrisTNG), finding that barred galaxies are less star forming and more gas poor than unbarred galaxies. Following the evolution of barred galaxies back in time, the authors find that these objects assembled most of their disc components and black holes before bar formation, and did this earlier than unbarred galaxies. In addition, \citealt{peschken2019}, indicate that flyby interactions are a key mechanism in the formation of bars in high-mass disc galaxies in the Illustris simulation, whereas the presence of gas in the disc can inhibit the formation of tidally induced bars and weaken existing bars over time.

Regarding the evolution of the bar fraction, \cite{rosasguevara2022} (see also \citealt{zana2022}) found that this bar fraction is almost flat as a function of redshift, also finding high-redshift bars as early as $z=4$ for the TNG50 simulations and in reasonable agreement with the fraction of the local Universe. The authors also find that if there are observational biases, such as those caused by limited angular resolution, the bar fraction evolution decreases with increasing redshift, as seen in observations. Similarly, a decreasing trend was found by \cite{fragkoudi2021} using the Auriga simulations \citep{grand2017}. In contrast, \cite{reddish2022}, using the NewHorizon simulation with higher resolution but smaller volume ($10$ Mpc on side), found that the bar fraction is smaller than in observations but decreases with increasing redshift. The main reason for this was found to be that spiral galaxies are dominated by dark matter in the central parts, which do not fulfil the conditions of forming a bar.

The main goal of the present study  is to identify the role of the environment in the evolution of disc galaxies and their bar structures. To this end, we follow the evolution of barred and unbarred disc galaxies ($z=1$) up to $z=0$ and investigate the formation and dissolution of their bars. We also explore whether or not there is a connection between the environment in which a galaxy resides and the likelihood that a bar will form or dissolve therein. We make use of the TNG50 simulation \citep{pillepich2019,nelson2019b}, which is the highest-resolution simulation run of the TNG project, covering a $51.7$ Mpc region.

The paper is structured as follows. In section~\ref{sec:method},  we describe the simulation, our parent disc galaxy sample, and environment proxies. In section~\ref{sec:morpevol}, we present the morphological evolution of the disc galaxies up to $z=1$ and their relation with the environment. In section \ref{sec:barred galaxies},  we study the evolution of the barred galaxies and the fate of their bar structures. In section  \ref{sec:unbarred galaxies}, we investigate the evolution of unbarred galaxies and examine whether or not they form a bar. In section \ref{sec:barfraction}, we discuss the effects of environment on bar fraction.  Finally, in section~\ref{sec:summary},  we summarise our findings.


\section{Methodology}
\label{sec:method}
\subsection{TNG simulations}
\label{subsec:TNGsims}
The IllustrisTNG (The Next Generation) project\footnote{\citep{nelson2019a}; http://www.tng-project.org}  (\citealt{nelson2018,naiman2018,pillepich2018b,marinacci2018,springel2018})
includes three main cosmological, gravo-magneto-hydrodynamical simulations of galaxy formation with volumes ranging from $(50)^3$ to $(300)^3\,\cMpc^3$  with different spatial  and mass resolutions.
The IllustrisTNG simulations were performed with the moving-mesh \textsc{AREPO} code \citep{springel2010},
adopting the Planck cosmology parameters with constraints from \cite{planck2016}: $\Omega_\Lambda=0.6911$, $\Omega_{\rm m}=0.3089$, $\Omega_{\rm b}=0.0486$, $\sigma_8=0.8159$, $h=0.6774$, and $n_{s}=0.9667,$  where  $\Omega_\Lambda$, $\Omega_{\rm m}$, and  $\Omega_{\rm b}$ are the average densities of matter, dark energy, and baryonic matter in units of the critical density at $z=0$,  $\sigma_8$ is the square root of the linear variance,  $h$ is the Hubble parameter ($H_{0}\equiv h \,100 \rm km \, s^{-1}$), and $n_{s}$  is the scalar power-law index of the power spectrum of primordial adiabatic perturbations.

The subgrid physics of IllustrisTNG is based on its predecessor, Illustris \citep{vogelsberger2013,vogelsberger2014a,vogelsberger2014b,genel2014,nelson2015,sijacki2015}, with substantial modifications to star formation feedback (winds), the growth of supermassive black holes, active  galactic nucleus (AGN) feedback, and stellar evolution and chemical enrichment. A complete description of the improvements made to the subgrid physics and calibration process can be found in \cite{pillepich2018a} and \cite{weinberger2017}.   A summary of the improvements concerning Illustris is shown in Table~1 of \cite{pillepich2018a}.

In this work, we focus on the \TNGF~simulation \citep{pillepich2019,nelson2019b}, which is the highest-resolution simulation that is part of the TNG suite and, at the same time, provides a sufficiently large cosmological volume to study the statistical properties of galaxies at intermediate masses. The simulation evolves $2160^3$ dark matter particles and initial gas cells in a 51.7 comoving 1 Mpc region from $z=127$ down to $z=0$. The mass resolution is $4.5\times 10^5 \Msun$ for dark matter particles, whereas the mean gas mass resolution is $8.5\times10^4\Msun$. A comparable initial mass is passed down to stellar particles, which subsequently lose mass through stellar evolution.
The spatial resolution for collisionless particles (dark matter, stellar, and wind particles) is 575 comoving pc down to $z=1$, after which it remains constant at 288 pc in physical units  down to $z=0$. In the case of the gas component, the gravitational softening  is adaptive and based on the effective cell radius
down to a minimum value of 72 pc in physical units, which is imposed at all times.

Galaxies and their haloes are identified as bound substructures using a friends-of-friends \textsc{(FoF)} and then a \textsc{SUBFIND} algorithm \citep{springel2001} and are then tracked over time by the \textsc{Sublink} merger tree algorithm \citep{rodriguezgomez2015}.
Halo masses ($M_{200}$) are defined as  all matter within the radius $R_{200}$ for which
the inner mean density is $200$ times the critical density.
In each FoF halo, the `central’ galaxy (subhalo) is the first (most massive) subhalo of each FoF group. The remaining galaxies within the
FoF halo are its satellites.  The stellar mass of a galaxy is defined as all the stellar matter assigned to the subhalo.

\subsection{Parent disc galaxy sample and identifications of bars}
\label{sub:discsample}
We focused on a subsample of the galaxy evolution, corresponding to  $z=1$ disc galaxies contained in the catalogue of \citealt[][hereafter  \citetalias{rosasguevara2022}]{rosasguevara2022}. Such galaxies were selected to have a stellar mass of $\geq10^{10}\Msun$ (accounting for more than $10^5$ stellar particles) to ensure that the galaxies analysed are well resolved and are dominated by a disc component ($D/T>0.5$). To identify the disc and bulge components of the galaxies, we use the kinematic decomposition computed in \cite{genel2015}, which is based on \cite{marinacci2014} and \cite{abadi2003} and is consistent with the selection of the disc sample in \citealt[][hereafter  \citetalias{rosasguevara2020}]{rosasguevara2020} for the TNG100 simulation. Galaxies are first rotated such that the z-axis is located along the direction of the total angular momentum of the stellar component. For each stellar particle within $10 \times \rhalf$, where $\rhalf$ is the radius within which 50 per cent of the total stellar mass is contained, the circularity parameter, $\epsilon=J_z/J(E)$, is calculated. $J_{z}$ is the specific angular momentum of the particle around the symmetry axis, and $J(E)$ is the maximum specific angular momentum possible at the specific binding energy of each stellar particle.
The mass of the stellar disc comprises the stellar particles with $\epsilon\geq 0.7$, while the bulge mass is defined as twice the mass of the stellar particles, with a circularity parameter $\epsilon<0$.
The disc-to-total ($D/T$) mass ratio is defined as the ratio between the disc stellar mass and the stellar mass enclosed in $10\times r_{50,*}$. We define disc-dominated galaxies as those galaxies with $D/T\geq0.5$ at $z=1$.

In addition to the bulge and disc decomposition described above, we employ the kinematic decomposition \textsc{mordor} computed in  \cite{zana2022}\footnote{https://github.com/thanatom/mordor} to determine more specific galaxy components. The decomposition is based on the circularity ($\epsilon$) and binding energy ($E$) phase space, where a minimum in $E$ is identified for each galaxy, $E_{\rm cut}$. The following four components can be identified and are used throughout the paper:
\begin{itemize}
    \item \textbf{Classical bulge:}  defined as those stellar particles that exhibit the highest degree of binding, the  value of which depends on the $E_{\rm cut}$ of each galaxy, and exhibit counter-rotation characterised by a negative value of $\epsilon$ ($<0$).  Next, a distribution that is equal to the component of the distribution exhibiting positive circularity is chosen and allocated to the bulge using Monte Carlo sampling.
    \item \textbf{Thin/cold disc:} defined as those stellar particles that exhibit the highest degree of binding and are not assigned to the bulge but have positive values of $\epsilon$ ($\epsilon>0.7$).
    \item \textbf{Pseudobulge:} defined as the remaining stellar particles that exhibit a high degree of binding, but are not assigned to the bulge or the thin disc.\footnote{We note that this definition of pseudobulge is not necessary to link to the bar and the definition of observed pseudobulge.}
    \item \textbf{Thick/warm disc:} defined as those stellar particles that exhibit a lower degree of binding and are not assigned to the bulge or pseudobulge, but with positive values of $\epsilon$ ($\epsilon>0.7$).
\end{itemize}

\subsection{Identification of bars}
Non-axisymmetric structures are identified by Fourier decomposing the face-on stellar surface density  (e.g. \citealt{athanassoula2002,zana2018a,rosasguevara2020}). We focus on  $\Ato(R)$, which is defined by the ratio between the second and zero terms of the Fourier expansion and its phase $\Phi(R)$ (see equations 1 and 2 in \citetalias{rosasguevara2022}). Both quantities $\Ato(R)$ and  $\Phi(R)$ have previously been used to characterise a bar structure, where the bar strength is defined as the value of the peak of $\Ato(R)$, $\Amax$. The phase should be constant inside the bar. We define a constant phase by calculating the standard deviation ($\sigma$) of $\Phi(R)$, including each time a new cylinder shell, and imposing $\sigma\leq 0.1$.  We then define the bar extent ($\rbar$) as the maximum radius where $\sigma\leq 0.1$, and the A2 profile first dips at  $0.15$ or the minimum value of $\Ato(R)$.
As large values of  $\Ato(R)$  could also be due to transient events, such as mergers and interactions, we conservatively assume the bar is a long-lasting feature if the following three criteria are fulfilled.
\begin{enumerate}
\item The maximum of $\Ato$ plus $\Amax$ is greater than $0.2$.
\item $\rbar>r_{\rm min}$ where $r_{\rm min}=1.38\times\epsilon_{*,z}$ is a  minimum radius imposed and $\epsilon_{*,z}$ corresponds to the proper softening length for stellar particles. To guide the reader, $r_{\rm min}$ spans from $0.16$ to $0.4\,\kpc$,
\item The estimated age of the bar is greater than the time between the analysed output and the two previous simulation outputs ($0.33$ Gyr at $z=0$ and $0.17$ Gyr at $z=4$).
\end{enumerate}

At lower redshifts, this latter criterion may exclude recently produced bars but remains consistent with the bar-selection criteria used by \citetalias{rosasguevara2020} when the presence of a bar structure was observed in earlier outputs.

In addition to the criteria used to identify a bar structure, we need to determine the formation time of a bar, $\tbar$. To this end, we retrace the evolution of $\Amax$ up to the lookback time when $\Amax \leq 0.2$  and  $\rbar(\tbar)\leq r_{\rm min}$  for more than two snapshots.  In addition, the relative difference between the $\Amax$ at a given snapshot and the $\Amax$ at the two prior snapshots must not exceed $0.45$ within this time frame. This provides additional assurance that the bars we detect are stable structures.

Finally, once the bar forms, the dissolution time of a bar $t_{\rm db}$ is determined by the lookback time when  $\Amax$ is smaller than $0.2$ and $\rbar(t_{\rm db})\leq r_{\rm min}$. In Appendix \ref{append:disolution}, we present two cases where the bar is dissolved.

The catalogue of \citetalias{rosasguevara2022} consists of 260 disc galaxies at $z=1,$ of which $125$ ($0.48\pm0.03$\footnote{These are the binomials errors on the bar fractions, which are calculated using
the number of galaxies and bars as $\sigma =(f_{\rm bars}(1- f_{\rm bars})/n_{\rm discs})^{0.5}$} ) have a bar. In this work, we ignore the galaxies that merged with a larger galaxy.
This condition reduced our sample to $204$ disc galaxies at $z=1,$ including $93$ barred galaxies ($0.45\pm0.03$). The details of the total sample can be found in Table~\ref{table:morphology}. We follow the evolution of these galaxies from $z=1$ to $z=0$. Hereafter, we tag the $z=0$ sample as descendant galaxies.

\subsection{Descendant galaxy samples}
\label{sub:descendantsamples}
In order to ascertain the primary mechanisms influencing the disc galaxies at $z=1$, we employ the \textsc{Sublink} merger tree technique \citep{rodriguezgomez2015} to track their evolution over time.  The parent disc sample was partitioned into three distinct subsamples based on the final morphology of their descendant galaxies at $z=0$.
Our analysis specifically centres on the stellar mass fraction of the cold-disc component at $z=0$ ($\Dthinzo$), as our initial dataset mostly consists of galaxies with a prominent cold-disc component.
The division was made in the following manner:
\begin{itemize}
    \item \textbf{Disc descendants} refer to those galaxies at $z=1$ and their descendant galaxies at $z=0$ that have  a  massive cold-disc component, defined as   $\Dthinzo\geq 0.5$ and $\Dthinzu\geq 0.5$.

    \item \textbf{Lenticular descendants} are defined as those galaxies at $z=1$ that have a  massive cold-disc component $\Dthinzu\geq 0.5$. These galaxies have descendants at $z=0$ that possess an intermediate cold-disc component, specifically falling within the range of $0.3\geq \Dthinzo< 0.5$.

    \item \textbf{Spheroidal descendants} defined as those galaxies at $z=1$ that have a  massive cold-disc component $\Dthinzu\geq 0.5$. However, their descendants at $z=0$ have seen a reduction in their cold-disc component, with values smaller than $\Dthinzo=0.3$.

\end{itemize}

The disc descendants comprise $104$ galaxies representing $51$ per cent of the total parent sample, of which  36 galaxies are barred at $z=1$. Among these 104 galaxies,  $47$ are classified as lenticular descendants ($23$ per cent of the parent samples) with $26$ barred galaxies at $z=1$. Finally, we get $53$ galaxies classified as spheroidal descendants, corresponding to $26$ per cent of the parent sample. The details can be seen in Table~\ref{table:morphology}. In the evolution of the bar structures, we only focus on disc and lenticular galaxies. 


\subsection{Quantifying the effects of environment}
\label{subsec:env}

The main aim of this study is to conduct a quantitative analysis of the impact of environmental factors on the development and progression of bars. Previous studies have provided evidence that mergers could facilitate the formation of a bar structure \citep[e.g.][]{elmegreen1990,lokas2018}. On the other hand, alternative studies have demonstrated that mergers can lead to the disruption of pre-existing bars \citep[e.g.][]{zana2018b}.
Tidal interactions, such as those resulting from galaxy--galaxy interactions and flybys, are an additional external mechanism that can contribute to the formation of bars \citep[e.g.][]{elmegreen1990,lang2014,lokas2018,peschken2019,izquierdo2022}. 
The aim of the present work is to quantitatively assess three environmental factors ---including the number of satellites as a function of time--- that may potentially influence the evolution and production of bars. By identifying these external mechanisms, we seek to gain a deeper understanding of their impact on bars.

\subsubsection{Merger histories and their influence}

Major mergers are characterised by a stellar mass ratio of the secondary galaxy to the primary galaxy, denoted $\mu$, that surpasses $0.25$, whereas minor mergers are defined by $\mu$ values ranging from $0.01$ to $0.25$ \footnote{The lower limit on the mass ratio for minor mergers ensures that both galaxies are properly followed. As our galaxy sample has stellar masses of $>10^{10}\Msun$, the secondary galaxy will have $M_{*}>10^8\Msun$ , which corresponds to  above $1000$ times the initial mass of the gas cells.}. In order to quantify the parameter $\mu$, the maximum stellar masses of the galaxies in preceding snapshots are used, namely those snapshots in which both galaxies are recognised as distinct structures by the \textsc{SUBLINK} algorithm.

According to \cite{McAlpine2020}, the period where mergers can strongly affect a galaxy can be defined as the period during which galaxies are expected to undergo a close merger within two dynamical timescales of the halo relative to the desired lookback time (time of bar formation, time of bar dissolution).  In this particular scenario, our focus lies on the temporal interval during which the bar undergoes formation or dissolution. The time period influenced by the merging process can be parameterised as
\begin{equation}
  \eta_{\rm dyn,obs} =  (t_{\rm m}- t_{\rm obs})/t_{\rm dyn,obs},
\label{eq:etadyn}
\end{equation}
where $t_{\rm obs}$ is the lookback time of formation or dissolution of the bar, $t_{\rm m}$ is the lookback time of the merger, and the dynamical time of the halo is defined as the free-fall time of a dark matter halo, that is,
\begin{equation}
  t_{\rm dyn,obs}\equiv \left( \frac{ 3\pi}{32G(200 \rho_{\rm crit})}\right)^{1/2}
.\end{equation}
This dynamical time of the halo depends on redshift. As a point of reference, the dynamical time of a halo for three redshifts is approximately $0.9$ Gyr at $z=1$, $1.2$ Gyr at $z=0.5$, and $1.6$ Gyr at $z=0$. Positive values of $\eta_{\rm dyn,obs}$ indicate that the closest merger has already occurred,  whereas negative values indicate that it will occur in the future.  Absolute values of $|\eta_{\rm dyn,obs}|<1$ indicate that the merger takes place in one dynamical time.
Therefore, we define the merger influence region for galaxies where the nearest merger occurs as $|\eta_{\rm dyn,obs}|<2,$ whereas values higher than $|\eta_{\rm dyn,obs}|>2$ are not influenced by a merger. Here, we focus mainly on major mergers, as they have the greatest potential for significant impact. However, we also calculated this for minor mergers.

\subsubsection{Close massive companions}

In a similar vein, to measure the likelihood that a galaxy will encounter a flyby event (not necessarily leading to a merger event), it is necessary to compute the proximity of the nearest large companion, where the ratio of their stellar masses exceeds $1/4$. Galaxies that may experience an influence due to a nearby encounter will possess nearby massive partners situated within a maximum distance of $d_{\rm cp}<100$ kpcs \citep{lokas2021}. Conversely, galaxies that are usually isolated are characterised by the closest massive companion being located at a distance of greater than $d_{\rm cp}>500$ kpc.





\section{Morphological evolution of $z=1$ disc galaxies}
\label{sec:morpevol}

\begin{figure}
\includegraphics[width=1\columnwidth]{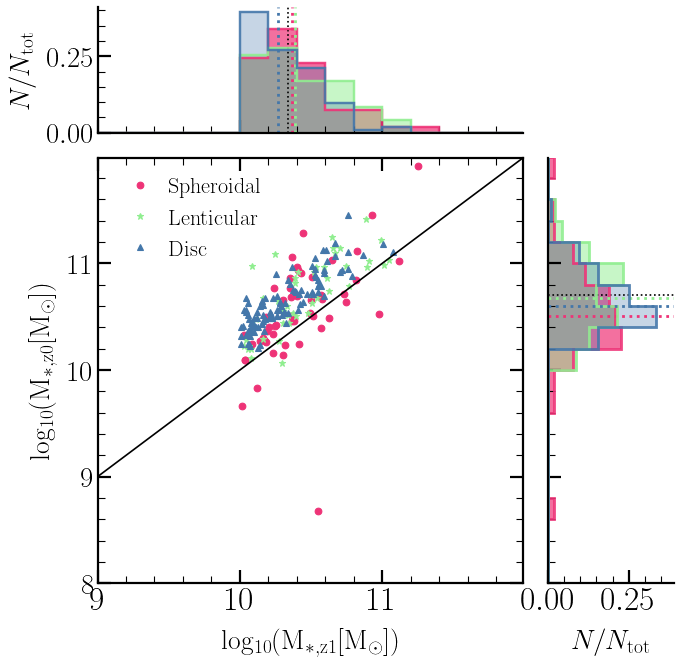}
\caption{Stellar mass of their descendants at $z=0$ as a function of the stellar mass of disc galaxies at $z=1$. Different markers and colours represent different morphologies of the descendant galaxies, as indicated in the legend. The black solid line represents a one-to-one relation. Panels along the margins show the distributions of stellar mass at $z=0$ and  $z=1$. Dotted lines represent the median of each distribution and the black dotted line is the median stellar mass of the total sample. There is a mild evolution in stellar mass in the last $8$ Gyr in the parent disc sample.}
\label{fig:massdistribution}
\end{figure}

Before analysing the evolution of the bar structures, we focus on the morphological evolution of our parent sample, as defined in section \ref{sub:discsample}.
We investigated the evolution of our parent disc galaxies at $z=1$ by employing the tree merger algorithm outlined in section \ref{subsec:TNGsims}. The primary focus of our analysis is on the galaxy descendants of these disc galaxies at $z=0$.

Figure~\ref{fig:massdistribution} displays the scatter plot of the stellar mass of the parent disc sample at $z=1$ versus the stellar mass of their descendants at $z=0$.  We can appreciate from the figure that, overall, there was an increase in stellar mass, although this growth is modest, with a median stellar mass of $10^{10.33}\Msun$ at $z=1$ rising to $10^{10.60}\Msun$ at $z=0$. We also note that some outliers, especially from spheroidal descendant galaxies, deviate from the overall distribution. The stellar mass of these galaxies decreases due to the entrance into a densely populated region, where they probably experience tidal stripping, as we show in the following section. This is confirmed when we look at the stellar-mass distributions (sided small panels) of the descendant galaxies split by shape, where spheroidal descendant galaxies have a smaller median stellar mass at $z=0$ than disc and lenticular descendant galaxies, whereas the parent sample at $z=1$ shows that the progenitors of the lenticular and spheroidal galaxies were more massive.



\begin{figure}
\includegraphics[width=1\columnwidth]{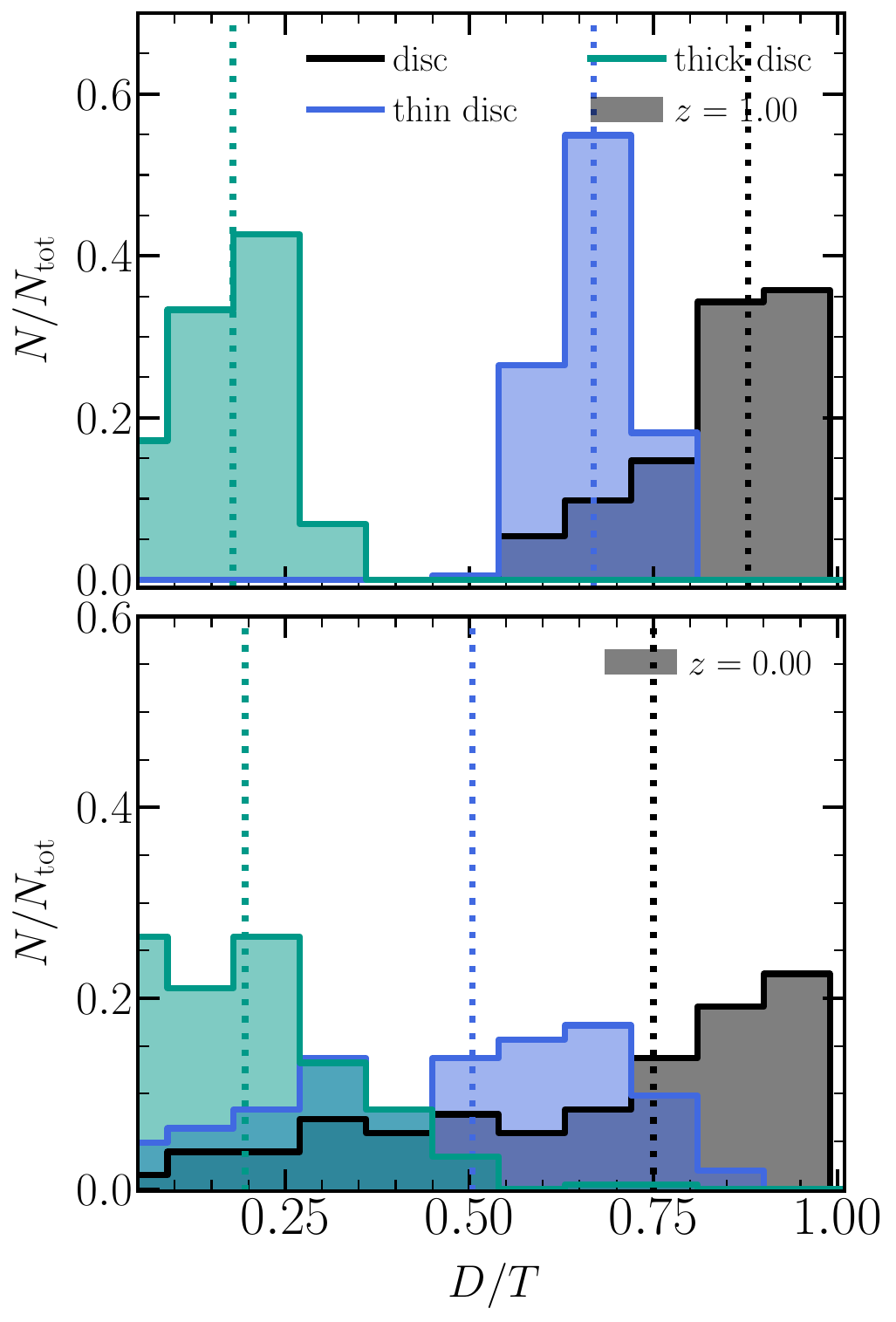}
\caption{ $D/T$ distribution of disc galaxies at $z=1$ (top panel) and their descendants at $z=0$ (bottom panel). We note that there was a morphological transformation overall in the population of the descendants of parent disc galaxies, especially for the thin disc component.}
\label{fig:distributionDtoTmordor}
\end{figure}

It is important to note that the morphological evolution of our parent sample exhibits significant divergence. Figure \ref{fig:distributionDtoTmordor} displays the distribution in the disc stellar mass fraction, denoted $D/T$ and represented in grey. The median $D/T$ decreases from $0.88$ at redshift $z=1$ to $0.75$ at redshift $z=0$. The dominant contribution is attributed to the thin-disc component, which exhibits a median of $\Dthinzu=0.67$ at $z=1$. Nevertheless, when considering the case where $z=0$, it can be shown that the distribution of $\Dthinzo$ in the cold-disc component is wider, spanning from $0.0$ to values higher than $0.75$. The median value of $(D/T)_{\rm thin,z0}$, is found to be $0.50$. This wider distribution of the cold component is also evident in the overall distribution of the discs, including all components. It should be noted that there is a lack of substantial evolutionary changes observed in the thick disc component, as indicated by a median value of $(D/T)_{\rm thin,z0,z1}\approx 0.2$ for both redshifts.

\begin{figure*}
\includegraphics[width=2\columnwidth]{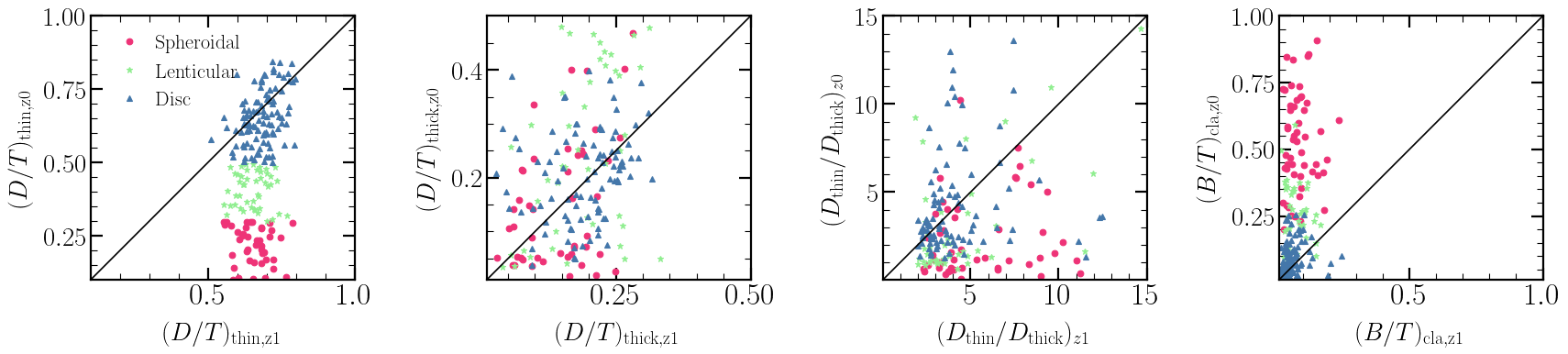}
\caption{Evolution of the cold and thin disc, warm and thick disc, and classical bulge components from $z=1$  for the three descendant subsamples: disc galaxies, lenticular galaxies, and spheroidal descendants.} 
\label{fig:evcomponents}
\end{figure*}

In order to examine the changes in morphology within our parent sample, we have chosen three galaxy descendant samples as outlined in section \ref{sub:descendantsamples} and Table \ref{table:morphology}. These samples have been categorised based on the prevalence of thin-disc components, as we can see in the left panel of Fig.~\ref{fig:evcomponents} where we can appreciate the stellar mass fraction of the cold disc at $z=0$ against that at $z=1$. The figure also shows a small evolution in the thick disc for the different samples, except for some lenticulars that show an increase by a factor or two.  When we compare the mass ratio between the thin disc and the thick disc, we see that this ratio increases with time for discs, whereas for lenticular and spheroidal galaxies, this ratio decreases, mainly because of the disruption of the cold disc.  Finally, the right panel of Fig.~\ref{fig:evcomponents}  shows the mass fraction of the classical bulge component for spheroidal descendants; this increases with time, reaching values of close to $1,$ whereas for disc galaxies, it increases up to $0.25$.
Based on the analysis of the descendant samples, it is observed that $51$ per cent of disc galaxies at redshift $z=1$ maintain their cold-disc component at a redshift of $z=0$, which accounts for $30$ per cent of the disc galaxies in the local Universe if we take the number of discs at $z=0$ (349) of \cite*{rosasguevara2022} for the TNG50. Additionally, we find that $26$ per cent of the $z=1$ disc galaxies transform into spheroidal galaxies, while the remaining $23$ per cent transition into lenticular galaxies (see Table~\ref{table:morphology}).

We also observe that a large fraction of disc galaxies ($49$ per cent) at redshift $z=1$ undergo a morphological transition. This transformation is likely attributed to external factors, such as tidal interactions or stripping resulting from mergers. Further details on this matter are discussed in the following section. Consequently, this phenomenon will have an impact on the pre-existing bar structures. Regarding disc descendants, we observe that a subset of the discs have the ability to develop or dissolve their bar.

\begingroup
\renewcommand{\arraystretch}{1.5}
\begin{table*}

\caption{Descendant galaxy samples defined according to their final morphology ($z=0$). From left to right:  Sample name, thin-disc mass fraction $\Dthin$ at $z=1$ ($\Dthinzu$), $\Dthin$ at $z=0$ ($\Dthinzo$), number of galaxies, the fraction with respect to the parent disc sample at $z=1$ ($f_{\rm gal}$), number of bars at $z=1$ ($n_{\rm bar,z1}$), number of bars at $z=0$ ($n_{\rm bar,z0}$), number of bars formed after $z=1$  $n_{\rm bar,form,z<1}$, number of dissolved bars ($n_{\rm bar,lost,z<1}$), number of unbarred galaxies ($n_{\rm{unbar}, z=1}$), and number of unbarred galaxies at $z=0$ ($n_{\rm{unbar}, z=0}$).   }
\centering
\begin{tabular}{lllllllllll} 
\hline
\hline
Name                         &  $\Dthinzu$ &   $\Dthinzo$       & $n_{\rm gal}$  & $f_{\rm gal}$  & $n_{\rm bar,z1}$ & $n_{\rm bar,z0}$ & $n_{\rm bar,form,z<1}$ & $n_{\rm bar,lost,z<1}$ & $n_{\rm{unbar}, z1}$ & $n_{\rm{unbar}, z0}$ \\
\hline
parent sample      &   $0.67$   &  $0.18$              &  $204$        &  $1$     & $93$&  $115$  & $59$          & $6$ & $111$    &      $36$
\\
disc descendants       & $\geq 0.5$ & $\geq 0.5$           &  $104$        & $0.51$         & $36$&  $74$   & $43$           & $5$ & $68$ & $30$
\\
lenticular descendants &  $\geq 0.5$ &  $\geq 0.3$, $<0.5$ &  $47$        &  $0.23$         & $26$&  $41$   & $16$          & $1$ &  $22$ & $6$ \\ 
spheroidal descendant &  $\geq 0.5$ &  $< 0.3$            &  $53$        &  $0.26$          & $31$&  $-$    & $-$           &  $-$ & $21$ & $-$ \\
\\
\hline
\end{tabular}
\label{table:morphology}
\end{table*}
\endgroup

\subsection{Merger histories and satellite fraction}

In this subsection, we examine the merger histories of the three descendant samples (disc, lenticular, and spheroidal) in which mergers have been identified, as discussed in section \ref{subsec:env}.

\begin{figure*}
\includegraphics[width=2\columnwidth]{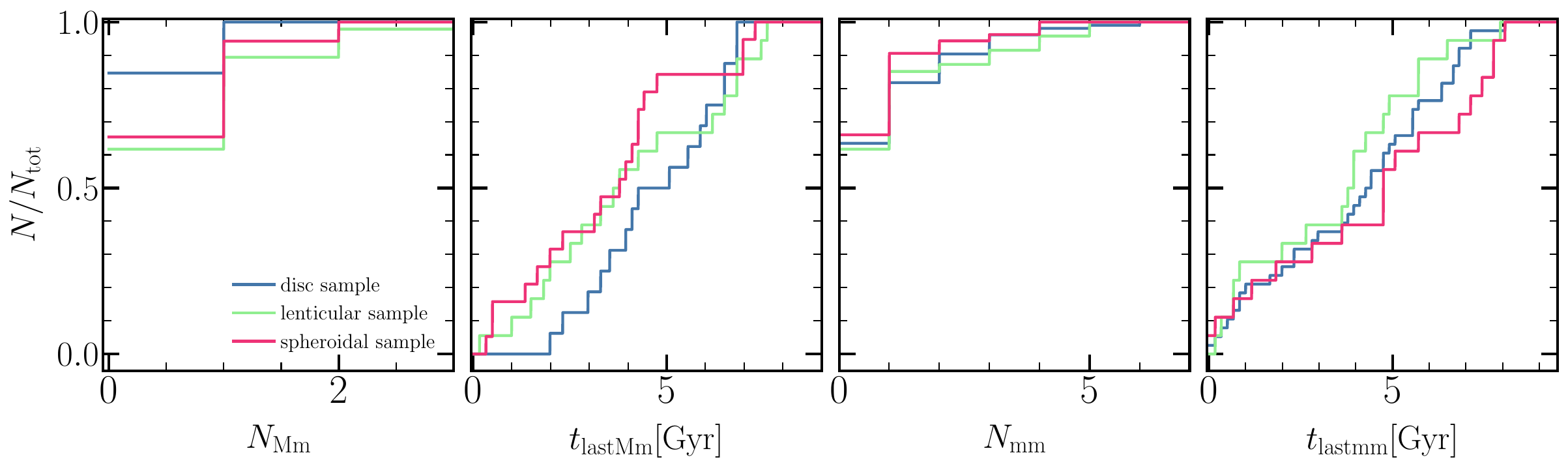}
\caption{Merger histories of the three galaxy samples as specified in the legend. From left to right: Cumulative distribution of major mergers, cumulative distribution of the last major merger, cumulative distribution of the minor mergers, and cumulative distribution of the last minor merger. Overall, the descendant disc galaxies have more quiet major merger histories when compared to lenticular and spheroidal descendant galaxies.}
\label{fig:mergerhistory}
\end{figure*}

We examine the cumulative distribution of major and minor mergers experienced by galaxies from redshift $z=1$ (corresponding to approximately 8 Gyr ago) to redshift $z=0$. The analysis is conducted on the descendant galaxy samples as depicted in Fig~\ref{fig:mergerhistory}. Overall, descendant galaxies with a disc morphology have relatively quiet merger histories. Specifically, $85$ per cent of these galaxies have not undergone a major merger in the past $8$ Gyr. In contrast, only $38$ per cent and $36$ per cent of the descendant lenticular and spheroidal galaxies, respectively, are free of major mergers. This is in agreement with previous theoretical studies (e.g. see Fig.16 in \citealt{izquierdo2019}). Additionally, in the second panel of Fig. \ref{fig:mergerhistory}, we see that disc descendant galaxies that have undergone a significant merger, on average, experienced their last major merger event approximately $4.7$ Gyr ago, but lenticular and spheroidal descendant galaxies typically underwent their last large merger later on, at  $3.7$ and $3.8$ Gyr ago, respectively.

Regarding the minor merger histories, no notable distinction is observed in the cumulative distribution across descendant galaxies of different types, with more than 30 per cent of them experiencing at least one minor merger. Specifically, approximately $62$ per cent of disc and lenticular galaxies and $66$ per cent of spheroidal descendant galaxies do not undergo minor mergers. The lenticulars underwent their most recent minor merger on average $3.9$  Gyr ago. In contrast, the disc and lenticular descendant galaxies exhibited earlier occurrences of minor mergers, thereby confirming the notion of relatively quiet evolutionary paths for these types of galaxies.  The data presented in Table~\ref{tab:discmergers} illustrate the fraction of galaxies that have undergone at least one significant merger event since redshift $z=1$ across various samples. Additionally, the table provides the median values for the most recent major and minor merger events within each sample.

\begingroup
\renewcommand{\arraystretch}{1.5}
\begin{table*}
    \caption{From left to right columns: Fraction of galaxies with at least one major merger since $z=1$, fraction of galaxies with at least one minor merger, and median lookback time of last major ($t_{\rm Mm}$[Gyr]) or  minor ($t_{\rm mm}$[Gyr]) merger.}
    \label{tab:discmergers}
    \centering
    \begin{tabular}{ccccc}
    \hline
                &$N_{\rm Mm}/N_{\rm tot}$&$N_{\rm mm}/N_{\rm tot}$ & $t_{\rm Mm}(N_{\rm Mm}>1) [\rm Gyr]$ & $t_{\rm mm}(N_{\rm mm}>1) [\rm Gyr]$\\
    \hline
   disc descendants  &  $0.15$   &  $0.38$ & $4.7$ & $4.3$  \\
   \hline
   lenticular descendants  & $0.38$   &  $0.38$  & $3.7$ & $3.9$ \\
   \hline
spheroidal descendants  &  $0.36$   &  $0.34$ &  $3.8$ & $4.7$  \\
   \hline
    \end{tabular}
\end{table*}
\endgroup

One significant element that may be playing a role in this morphological transformation is the influence of tidal torques generated by the dense environment in which these galaxies are located \citep[e.g.][]{martinez2016,lokas2016,lokas2021,mendez2023}.  This influence becomes particularly pronounced when these galaxies transition into denser environments.  One way to gain insight into this phenomenon is by examining the evolution of the satellite fraction for each descendant sample, because satellite galaxies are susceptible to experiencing heightened levels of tidal torque, ram pressure stripping, and mergers.

\begin{figure}
\includegraphics[width=1\columnwidth]{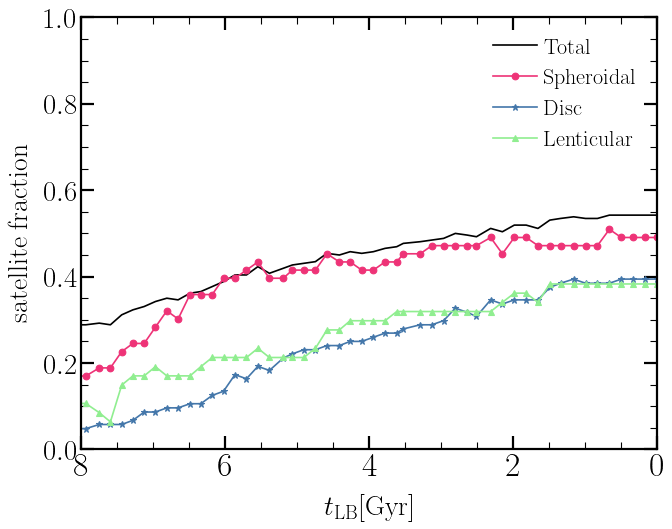}
\caption{Redshift evolution of the satellite fraction for the parent sample, the disc sample, and the lenticular and spheroidal descendant samples. The satellite fraction increases with decreasing redshift, with the highest evolution presented for the spheroidal descendants.}
\label{fig:satellitefraction}
\end{figure}

Figure \ref{fig:satellitefraction} displays the fraction of satellites in the parent sample and each descendant sample. Overall, the satellite fraction increases with time. The highest increase is presented by the spheroidal descendants, with satellite galaxies accounting for $20$ per cent at redshift $z=1$ and expanding to $50$ per cent in the current epoch.  Disc and lenticular descendants also present an increase,  albeit more modest. Specifically, the fraction of galaxies classified as satellites is less than $10$ per cent at redshift $z=1$, and gradually increases to reach approximately 40 per cent.   
These galaxies are marked by a higher occurrence of interactions and a larger degree of exposure to tidal forces.
Several processes in denser environments have the capacity to either disrupt or create a bar structure. This topic is further examined in the following sections.




\section{Evolution of the $z=1$ barred galaxies}
\label{sec:barred galaxies}
In this section, we explore the evolution of the bar structures in the barred galaxies at $z=1$. For this, we focus on disc and lenticular galaxies.
As discussed in the previous section, disc descendant galaxies, despite having relatively quiet merger histories in general, exhibit a diversity of merger histories and environments (see Fig \ref{fig:mergerhistory}), which must be taken into account when investigating the evolution of bar structures. For the lenticular galaxies, this diversity in the merger histories is higher. Each barred galaxy in the disc and lenticular descendants has been divided into two subsamples: (1) the descendants of barred galaxies at $z=1$ remain as barred galaxies at $z=0$; and (2) the descendants of barred galaxies at $z=1$ lose their bar, becoming unbarred galaxies at $z=0$.


\begin{figure}
\includegraphics[width=1\columnwidth]{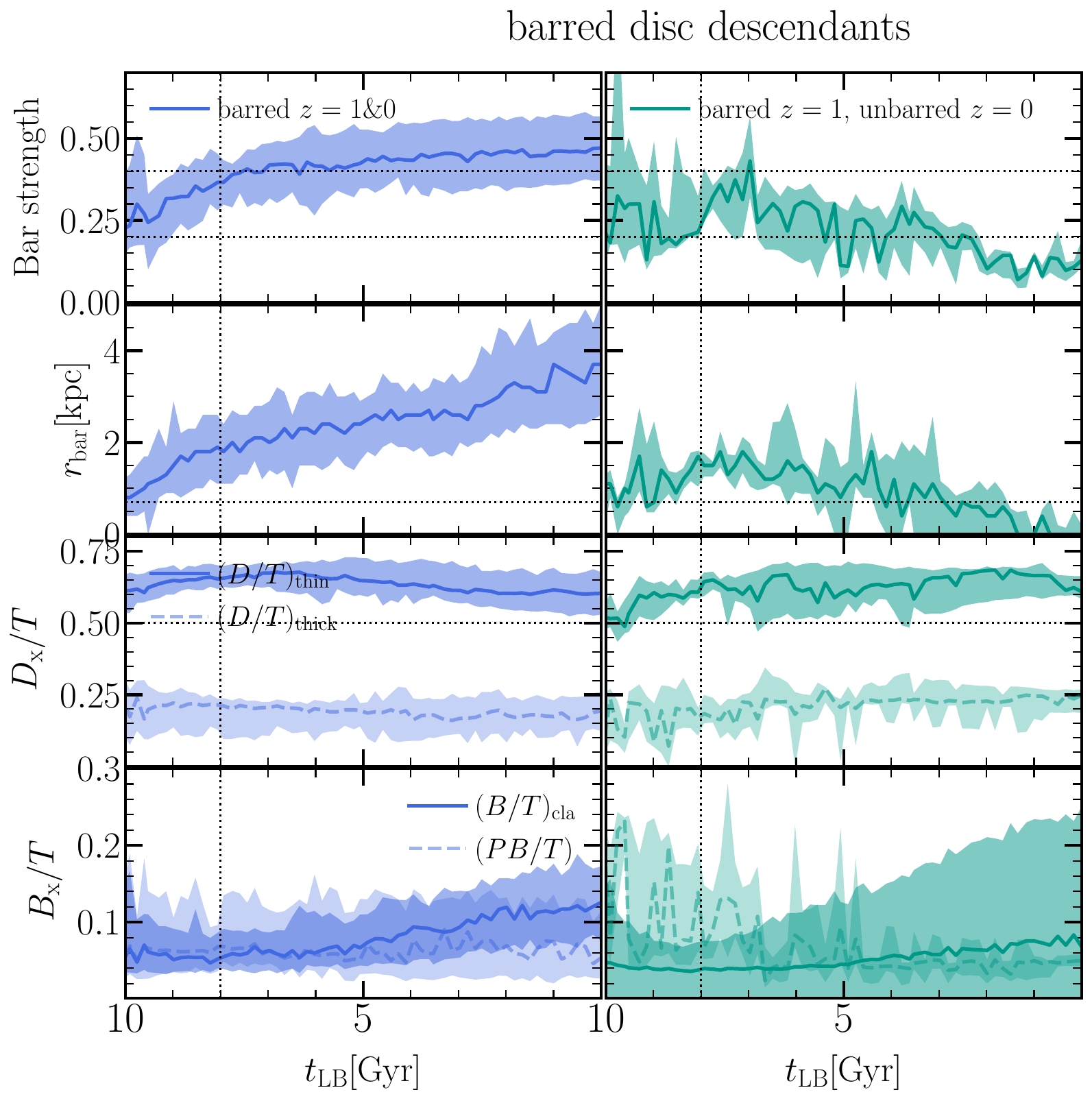}
\caption{Evolution of barred disc galaxies divided into two subsamples: barred galaxies at $z=1$ that remain barred galaxies at $z=0$ (left column), and those that do not have their bar at $z=0$ (right column). Solid lines represent the median values and the shaded area corresponds to the $20^{\rm th}$ and $80^{\rm th}$ percentiles of the distribution.}
\label{fig:evolutiondiscs}
\end{figure}

\begin{figure}
\includegraphics[width=1\columnwidth]{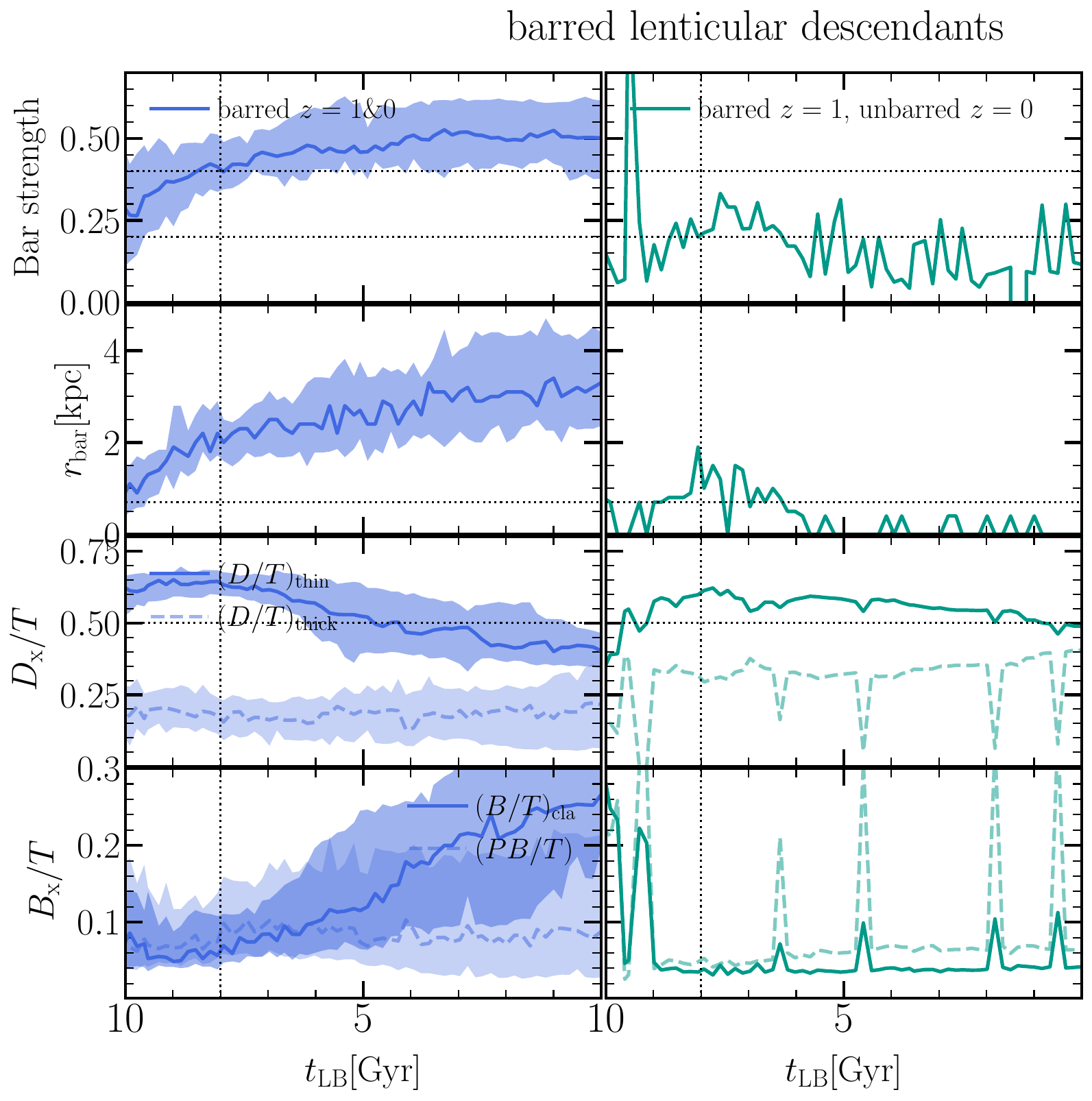}
\caption{Evolution of lenticular descendant galaxies divided into two subsamples: Barred galaxies at $z=1$ that remain barred  at $z=0$ (left column), and those that do not have their bar at $z=0$ (right column). The evolution of disc galaxies is shown in  Fig. \ref{fig:evolutiondiscs}. Solid lines represent the median values and the shaded area corresponds to the $20^{\rm th}$ and $80^{\rm th}$ percentiles of the distribution. We note that here we only have one lenticular galaxy that loses its bar.}
\label{fig:evolutionlenti}
\end{figure}
In our analysis, which is summarised in Table~\ref{table:morphology}, it is shown that a total of six galaxies, or $3$ per cent of the parent sample (the $z=1$ disc galaxies), lose the bar structure.  On the other hand, one-third of the galaxies in our parent sample, namely $56$ barred galaxies or $30$ per cent (including bars only in lenticular and disc galaxies), maintain a bar structure at $z=0$. Upon examination of each descendant barred galaxy sample, a consistent pattern emerges: a minority of disc and lenticular descendent galaxies experience bar dissolution, with fractions of $4.7$ per cent and $1.1$ per cent, respectively.

\subsection{Galaxy and bar properties across time}
The evolution of the properties of the bar structures is illustrated in Fig. \ref{fig:evolutiondiscs}  and Fig. \ref{fig:evolutionlenti} for the barred disc and lenticular galaxies, respectively.  The first columns of the figures depict the evolution of the bar structures in the descendant galaxies that have retained their bars until the present epoch ($z=0$). We can appreciate that the bars that were already formed at $z=1$ formed at an early time (approximately $10$ Gyr ago), and the evolution of their bars is typical of that observed in isolated galaxies. Previous studies \citep{martinez2006} showed that bars experience rapid growth during their dynamical phase, which lasts a few gigayears,  after which their evolution becomes secular and remains relatively constant.   However, there is a possibility that the bars may slightly weaken over time due to a buckling instability ---as explained by \cite{martinez2006}---, and then remain constant in their secular evolution.  This instability could potentially account for some of the fluctuations observed in the length of bars.   Upon examining the morphological evolution depicted in the third and fourth rows of  Fig. \ref{fig:evolutiondiscs} and Fig. \ref{fig:evolutionlenti},  it becomes evident that there is a lack of significant evolution in the thick-disc fraction. However, a marginal drop in the thin-disc fraction is observed (also seen in the evolution of the lenticular sample). This phenomenon is associated with the rise in the bulge mass fraction ($B/T$) of both classical bulges and pseudobulges during later times \cite[see also][]{izquierdo2022}. This increase can likely be attributed to the presence of a bar, which facilitates the inflow of gas towards the central regions of galaxies
\citep{spinoso2017,donohoe2019,george2019, fraser2020} or may be the result of mergers; although in the following section we  show that barred galaxies of the disc have the quietest merger histories. This increase in the $B/T$  is also found in lenticular galaxies, and the magnitude of this increase is greater. It is important to observe that the bars from the descendant barred disc galaxies that have retained their bar structure are more elongated than those in lenticular galaxies.

The evolution of the bar structures that are dissolved is shown in the right columns of Fig. \ref{fig:evolutiondiscs} and Fig. \ref{fig:evolutionlenti}. These galaxies under consideration constitute a smaller fraction, specifically $4.7$ and $1.1$ per cent, of the total population of disc and lenticular descendants, respectively. We notice that these bars are young and small, and just recently formed before $z=1$. Regarding morphology,  the disc mass fraction remains relatively stable at $z=0$ for the disc galaxy samples, with occasional changes likely attributed to either a major or a minor merger event. The bulge mass fraction exhibits a noticeable rise but with a significant degree of scatter.  Additionally, it is important to note that the thick disc and pseudobulge components do not exhibit any signs of evolutionary change.
In the case of the lenticular galaxies that lose their bar (only one case), not only is the bar structure disrupted but so is the thin disc,  whereas there is no increase in the classical and pseudobulge mass fractions, indicating that the galaxy seems to be disrupted by its environment.

\subsection{The role of environment in barred galaxies}
\label{sec:environ}
\begin{figure}
\includegraphics[width=1\columnwidth]{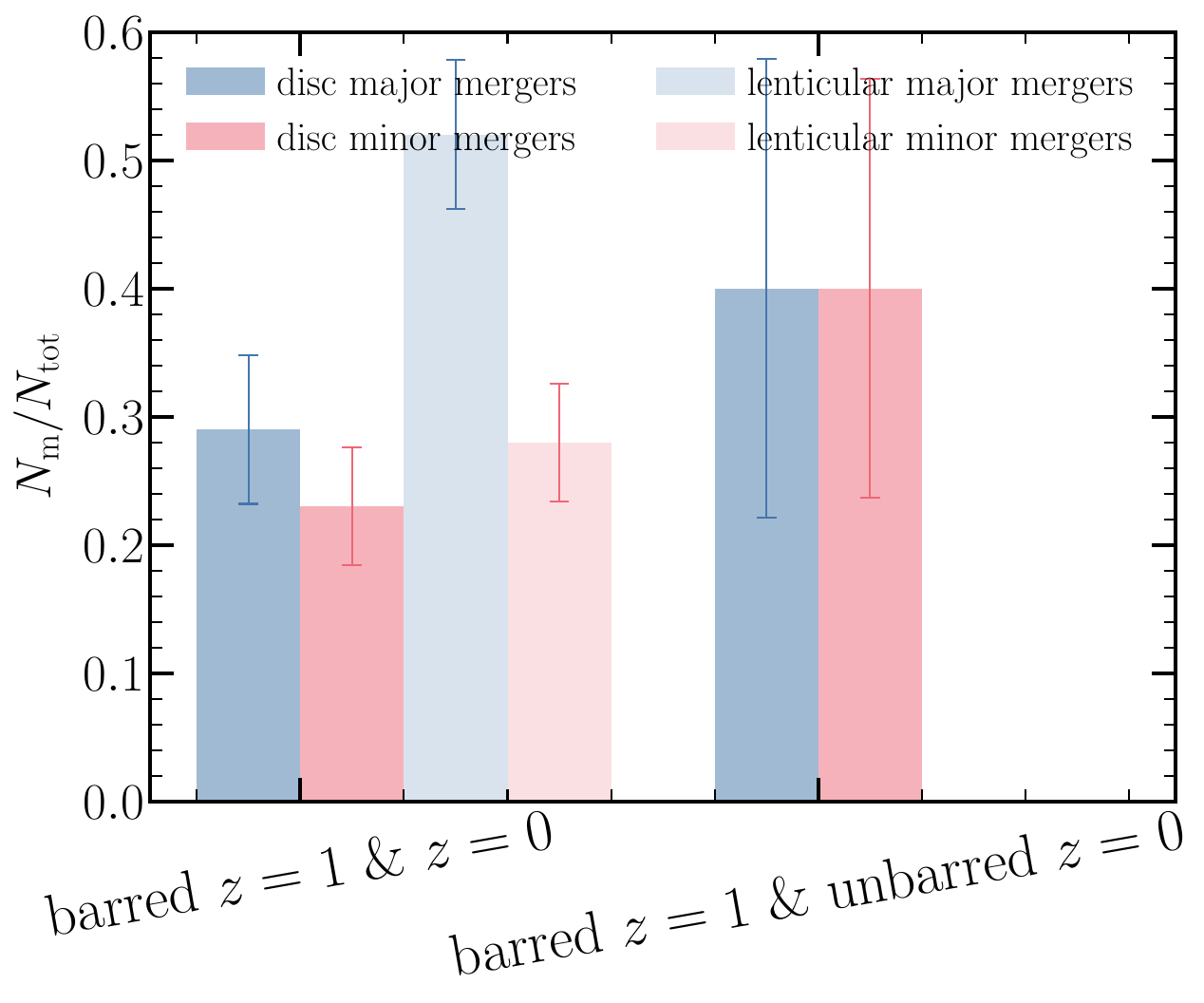}
\caption{Fraction of barred galaxies that experienced a major or minor merger since $z=1$ for the disc and lenticular descendants. Error bars represent the poison errors of each sample. The highest fraction of major or minor mergers corresponds to the subsamples where a bar structure dissolves, whereas the quiet histories are from disc galaxies that preserved their bar.}
\label{fig:mergersbar}
\end{figure}

In this section, we examine the effect of environment on the bar structures of disc and lenticular descendants of barred galaxies. To accomplish this, we first examine the merger histories of each sample.
To increase our statistics in this section, we analyse the lenticular and disc galaxies together.
Figure \ref{fig:mergersbar} illustrates the fraction of descendants that belong to the barred disc and lenticular categories and have experienced major or minor mergers since redshift $z=1$.  Our study shows that the galaxy fractions with the lowest values of major and minor mergers (less than $0.28$ and $0.25$, respectively) are seen for barred disc descendant galaxies that have maintained their bar structure.  The maximum fraction of galaxies experiencing major mergers is observed when disc galaxies undergo bar dissolution, with an approximate value of $0.40$. The fraction of disc galaxies that have experienced a minor merger exhibits a consistent trend across all subsamples. In particular, galaxies that are descendants of barred discs and have retained their bar structure show the lowest fractions.
Conversely, barred galaxies that lose their bar present the highest fractions.  It should be noted that the fraction of galaxies experiencing a minor merger is small in relation to those undergoing major mergers. Our findings indicate a correlation between the dissolution of a bar and the merger histories of galaxies. Indeed, previous studies proposed that mergers can also lead to the weakening of a bar, especially if they have a contribution to the central part of the galaxy, as the stronger gravitational potential can  weaken the bar \citep[e.g.][]{zana2018b,ghosh2021}.

Regarding lenticular galaxies, there is an observed trend that is in contrast to what is typically observed in disc galaxies. In particular, galaxies that have retained their bars are the most likely to have undergone a significant merger ($>0.50$), while galaxies that have not developed bars demonstrate a comparable ratio ($0.40$).  This increase may account for the observed rise in the mass percentage of the bulge and the disruption of the disc (as depicted in Fig \ref{fig:evolutionlenti}) during later stages of the evolution of barred lenticular descendant galaxies that have retained their bar structure. Because we have a smaller sample size of lenticular galaxies that have lost their bar, we do not have information about the impact of minor merges. This finding implies that the environmental factors surrounding the dissolution of a bar after redshift $z=1$ may play a significant role.


\begin{figure}

\includegraphics[width=1\columnwidth]{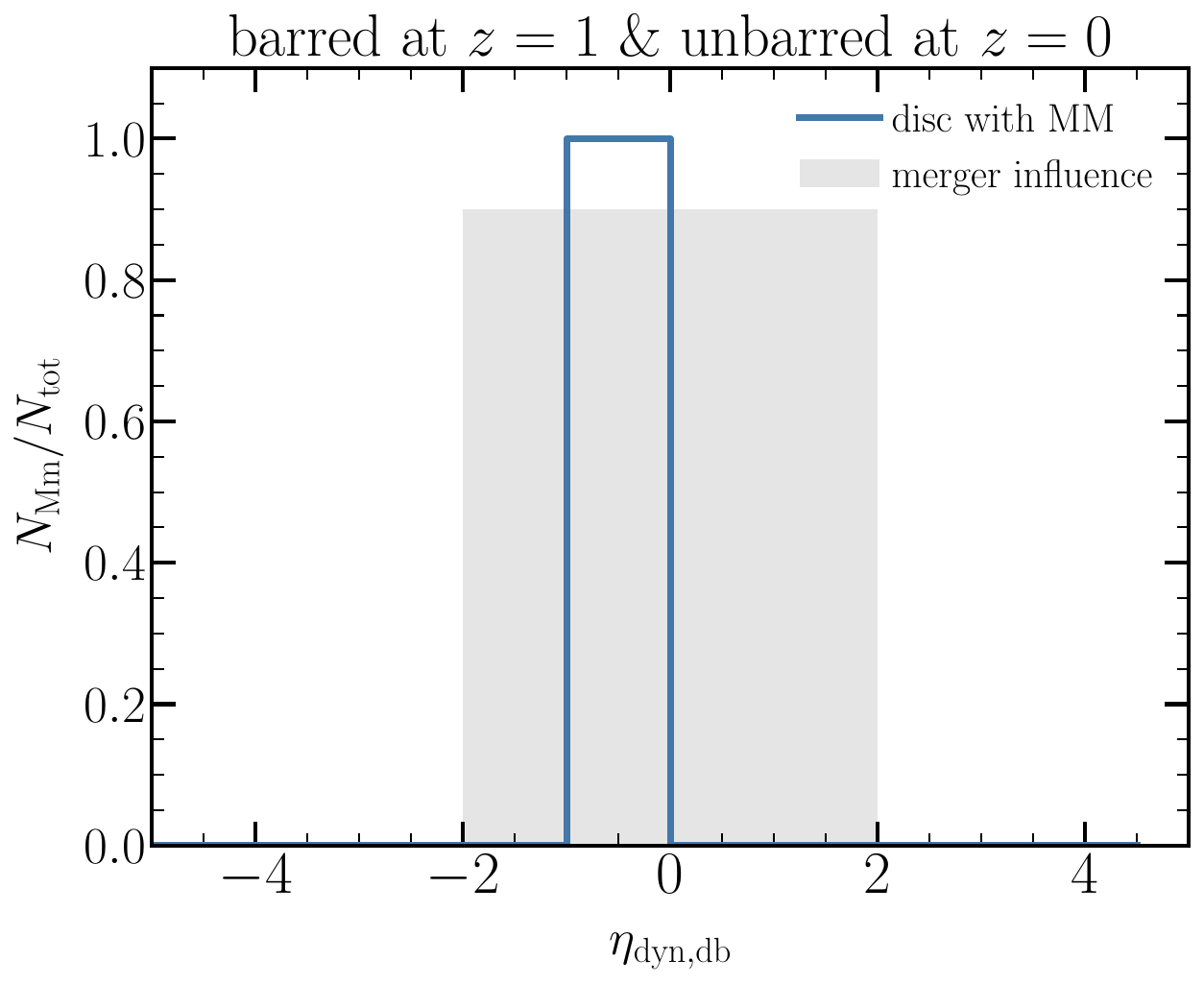}

\caption{Distributions of  $\eta_{\rm dyn,db}$ defined in Eq. \ref{eq:etadyn} for barred galaxies that lose their bar and undergo a major merger. This expression assesses whether or not  the disc galaxy experienced a recent merger at the moment of bar dissolution, and values between $-2$ and $2$ indicate that the galaxy is still influenced by a recent merger. Here, we see that 50 per cent of the disc galaxies that have lost their bar experienced a recent merger. Lenticular galaxies are not included because of the lower sample size we have for lenticular galaxies that have lost their bar.}
\label{fig:tdynbu}
\end{figure}

To closely examine the impact of major mergers in the galaxies that have retained their bars and compare them to those that have undergone bar dissolution, we calculated the temporal zone of merger influence, as stated in section \ref{subsec:env}. Quantification of this influence can be achieved through the parameter $\eta_{\rm dyn,db}$, as defined in  Eq. \ref{eq:etadyn}. With this parameter, the discrepancy between the time of dissolution and the time of the most recent major or minor merger or the upcoming major or minor merger is computed relative to the dynamical time of the halo at the moment of bar dissolution. As a gentle reminder to the reader, it is important to note that negative values of $\eta_{\rm dyn,db}$ indicate the occurrence of a future merger, whereas positive values indicate that a merger has occurred in the past. The temporal range within which the merger may have an impact on the galaxy, relative to the dynamical time of the halo, spans from $[-2,2]$.

To evaluate the impact of significant interactions on the disruption of a bar, we also incorporate the right panel of Fig \ref{fig:tdynbu}, which illustrates the distribution of  $\eta_{\rm dyn,db}$ values for galaxies that experience the loss of their bar and undergo a significant merger in close temporal proximity to the onset of bar weakening. As depicted in the diagram, it is observed that the barred disc galaxies undergoing a major merger experience a weakening of the bar, which finally leads to its disappearance.  The suggested scenario is not applicable to lenticular galaxies, for which we do not have an example. However, the weakening of the bar could be attributed to tidal interactions induced by flybys and this could potentially destroy them \citep[e.g.][]{zana2018b}.

\begin{figure}
\includegraphics[width=1\columnwidth]{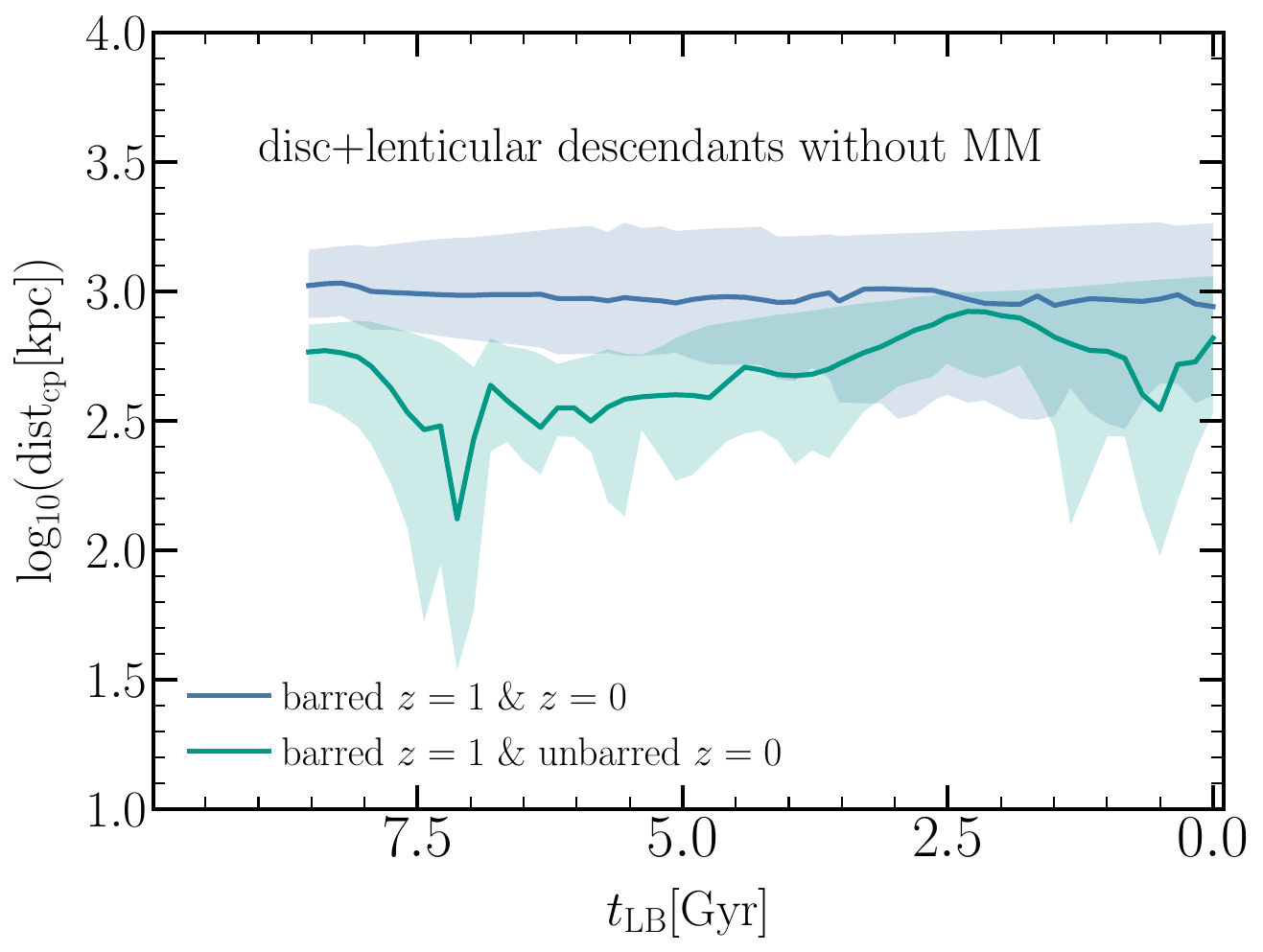}
\caption{Median distance to a massive companion (mass ratio $=1/4$) for disc and lenticular descendant galaxies that have lost their bar and have not undergone a major merger (green line) versus barred galaxies that have not undergone a major merger as a function of time since $z=1$. The shaded area corresponds to the $20^{\rm th}$ and $80^{\rm th}$ percentiles of the distribution. Galaxies that have lost their bars are closer to their nearest massive companion than galaxies that have retained their bars. }
\label{fig:closeneigh}
\end{figure}

\begin{figure}
\includegraphics[width=1\columnwidth]{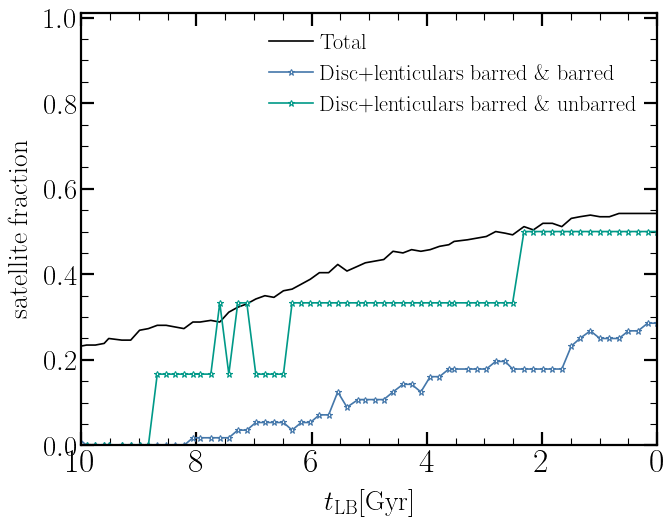}
\caption{Redshift evolution of the satellite fraction of barred discs and lenticular galaxies that retain their bar versus that of  barred discs and lenticular galaxies that lose it. In general, the satellite fraction of lenticular and discs that lose their bar is higher in comparison to those that retain their bar.}
\label{fig:fracsatellitetypebars}
\end{figure}

As a means of investigation, we computed the distance to the closest massive companions, as outlined in section \ref{subsec:env}, with the condition that the mass stellar ratio between the companion and the primary galaxy exceeds $1/4$. We performed this calculation for lenticular and disc galaxies that have undergone bar dissolution and have not experienced a significant merger event, and show the result as a function of lookback time.  This is shown in Fig. \ref{fig:closeneigh}, where it is evident that the galaxies being examined display diverse interactions with massive companions at distances of less than $100$ kpc, particularly during early stages, near $z=1$. In order to establish a basis for comparison, we further computed the distance to the closest massive companion for barred disc and lenticular galaxies that have retained their bars and have not experienced significant merger events. We find that the distance to a massive galaxy is systematically higher in comparison with galaxies that lose their bar. We note that this occurs for lenticular and disc descendants, suggesting that tidal interactions are an important factor in the dissolution of bars. More precisely, taking into account lenticular and disc galaxies, $50$ per cent of them become satellites before the bar weakens, with a difference between the time of dissolution and the time of becoming a satellite being less than $1$ Gyr, except for one case where this takes almost $2$ Gyr, including the galaxies with mergers.  This suggests that entry into a more dense environment contributes to the destruction of a bar. This is in agreement with a recent study of cluster galaxies from the extended WIde-field Nearby Galaxy-cluster Survey (OmegaWINGS), where the authors found that galaxies situated on the periphery of a galaxy cluster are susceptible to tidal interactions, which have the potential to cause the destruction of the bar \citep{tawfeek2022}.

We conclude this section by presenting the satellite fraction for disc and lenticular descendants in the last 8 Gyr ($z=1$). Figure \ref{fig:fracsatellitetypebars}  depicts the evolution of the satellite fraction for the barred discs and lenticular galaxies that retain their bar and compares this to the evolution of the satellite fraction of those that lose their bar. As expected, disc galaxies that lose their bar, in general, live in denser environments and then present a higher satellite fraction ($0.50$) in comparison to barred disc and lenticular galaxies that retain their bar ($0.30$). This happens later in time, which is in agreement with the idea that the dissolution of a bar could be triggered by tidal interactions by a close encounter or by ram pressure stripping due to torque forces in a dense environment. In conclusion, our results suggest that galaxies that lose their bar are exposed to tidal interactions.

\section{The evolution of $z=1$ unbarred galaxies}

\label{sec:unbarred galaxies}
In this section, we explore the evolution of the non-axisymmetry structures in the unbarred galaxies at $z=1$ to determine whether or not they develop a stable bar. For this, we focus on disc and lenticular descendants. As in the previous section, we break into two groups of unbarred galaxies: (1) the descendants of unbarred galaxies at $z=1$ that form a bar later in time, becoming barred galaxies at $z=0$; and (2) the descendants of unbarred galaxies at $z=1$ that were not able to develop a stable bar and remain unbarred at $z=0$.

We find that $59$ unbarred galaxies, accounting for $29$ per cent of the parent sample (the $z=1$ disc galaxies), develop a stable bar structure at any point between $z=1$ and $z=0$. On the other hand, we find that 36 disc and lenticular galaxies ($18$ per cent of the parent sample) do not possess a bar structure at redshift $z=0$ but  were able to develop a stable bar structure after $z=1$.

\begin{figure}
\includegraphics[width=1\columnwidth]{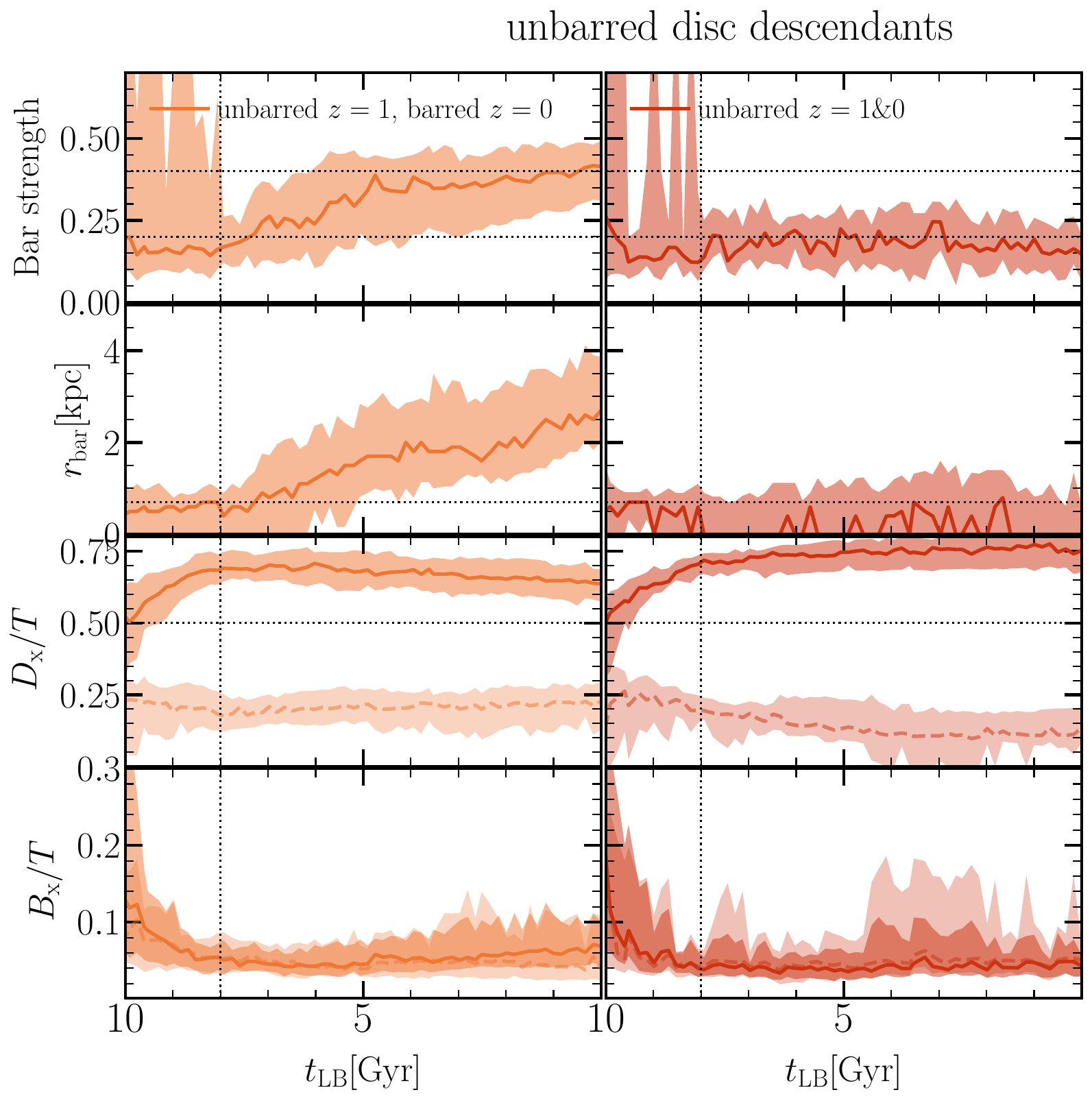}
\caption{Evolution of unbarred disc galaxies divided into two subsamples: unbarred galaxies at $z=1$ that become barred galaxies at $z=0$ (left column), and  those that were not able to develop a stable bar structure at any point up to $z=0$ (right column). Solid lines represent the median values and the shaded area corresponds to the $20^{\rm th}$ and $80^{\rm th}$ percentiles of the distribution.}
\label{fig:evolutiondiscsunbarred}
\end{figure}

\begin{figure}
\includegraphics[width=1\columnwidth]{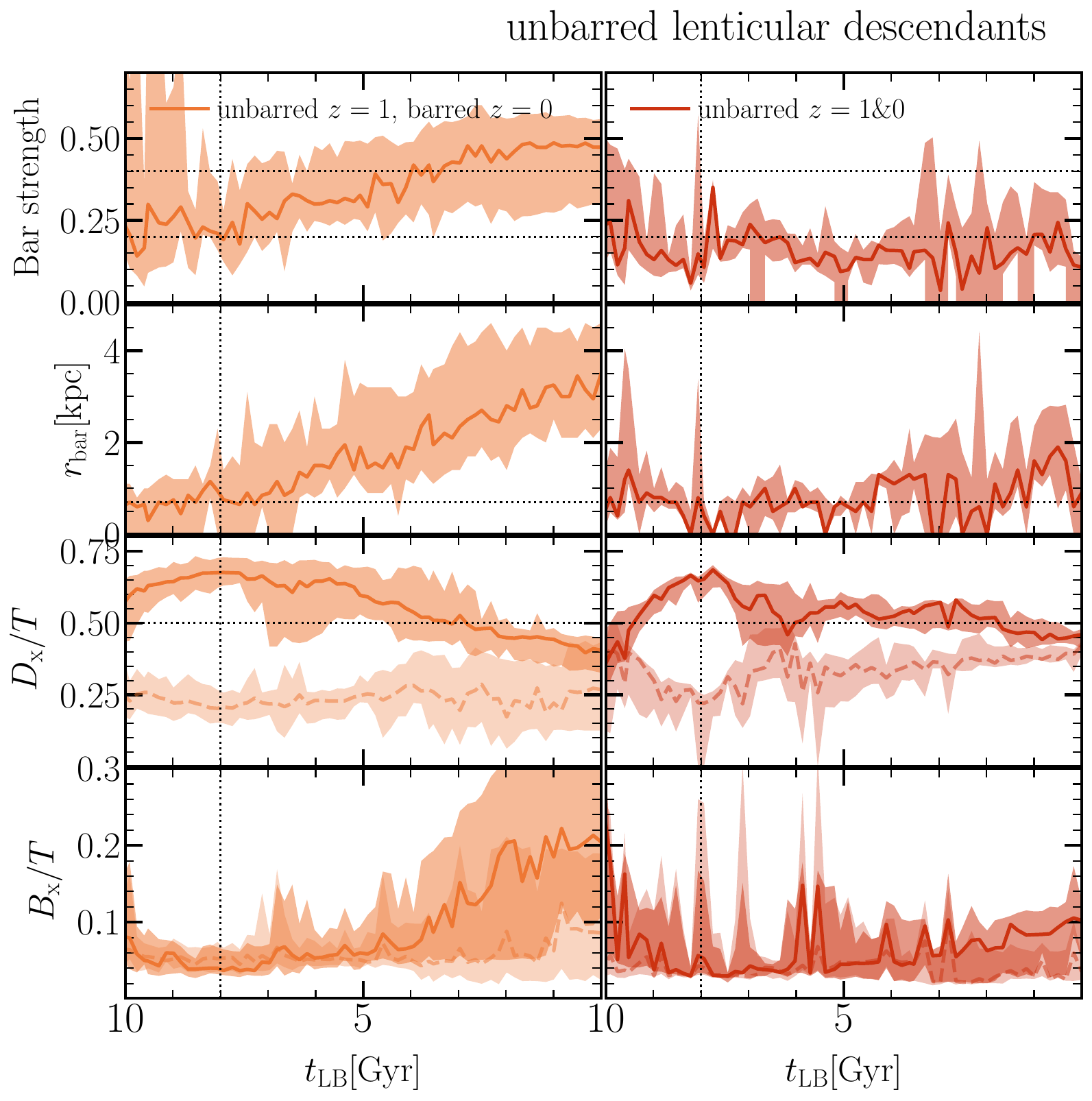}
\caption{Evolution of unbarred lenticular galaxies divided into two subsamples as in Fig.\ref{fig:evolutiondiscsunbarred}: unbarred galaxies at $z=1$ that became barred galaxies at $z=0$ (left column), and those that were not able to develop a stable bar structure at any point up to $z=0$ (right column). Solid lines represent the median values and the shaded area corresponds to the $20^{\rm th}$ and $80^{\rm th}$ percentiles of the distribution.}
\label{fig:evolutionlentiunbarred}
\end{figure}

The evolution of the non-axisymmetric structures is illustrated in Fig. \ref{fig:evolutiondiscsunbarred} over the disc descendants.  The first column depicts the evolution of the non-axisymmetric structures in the descendant galaxies that form a bar after $z=1$ and keep it until the present epoch ($z=0$). It can be seen that the formation of bar structures in these galaxies occurs shortly after the first few gigayears. It should be noted that, on average, the bar is comparatively weaker and smaller compared to the subsample of barred galaxies that have managed to retain their bar structure (see Fig.~\ref{fig:evolutiondiscs}), which is in line with them forming later in time. In terms of morphology, no significant evolutionary changes in the stellar mass fractions of the galaxy components are observed in this subsample.

Similarly, the bar structures in lenticular descendants are formed just after $z=1$ as seen in Fig. \ref{fig:evolutionlentiunbarred}. It should be noted that in the case of lenticular galaxies, the presence of a bar structure occurs during a stage where the progenitor galaxies possess a significant cold-disc component and a relatively small bulge component before the morphological transformation takes place, and on average, these bars are stronger and longer than counterparts hosted by disc galaxies.

The evolution of the $A_{2,\rm max}$ can be seen in the second column of Fig. \ref{fig:evolutiondiscsunbarred} for unbarred disc galaxies that never develop a stable bar. It is interesting that these galaxies have a relatively tiny bulge mass fraction and a correspondingly high disc mass fraction. Furthermore, there seems to be limited evolution observed in the thick disc and pseudobulge components. In the case of lenticular galaxies, there is an increase in the thick-disc fraction and a decrease in the thin-disc fraction, whereas there is no or little evolution in the pseudobulge and bulge components. This is shown in the left panel of Fig. \ref{fig:evolutionlentiunbarred}.

To understand why some unbarred galaxies develop a bar whereas there are others that do not, we investigated the role of the environment. We present our findings in the following section.

\subsection{The role of environment in unbarred galaxies}
\label{subsec:unbarredmergers}
Previous studies proposed that mergers and tidal interactions can be another potential formation channel of bars \citep[e.g.][]{lokas2016,martinez2016,lokas2018,izquierdo2022}. In this section, we investigate whether or not environment has any influence on the likelihood that an unbarred galaxy will form a bar.

\begin{figure}
\includegraphics[width=1\columnwidth]{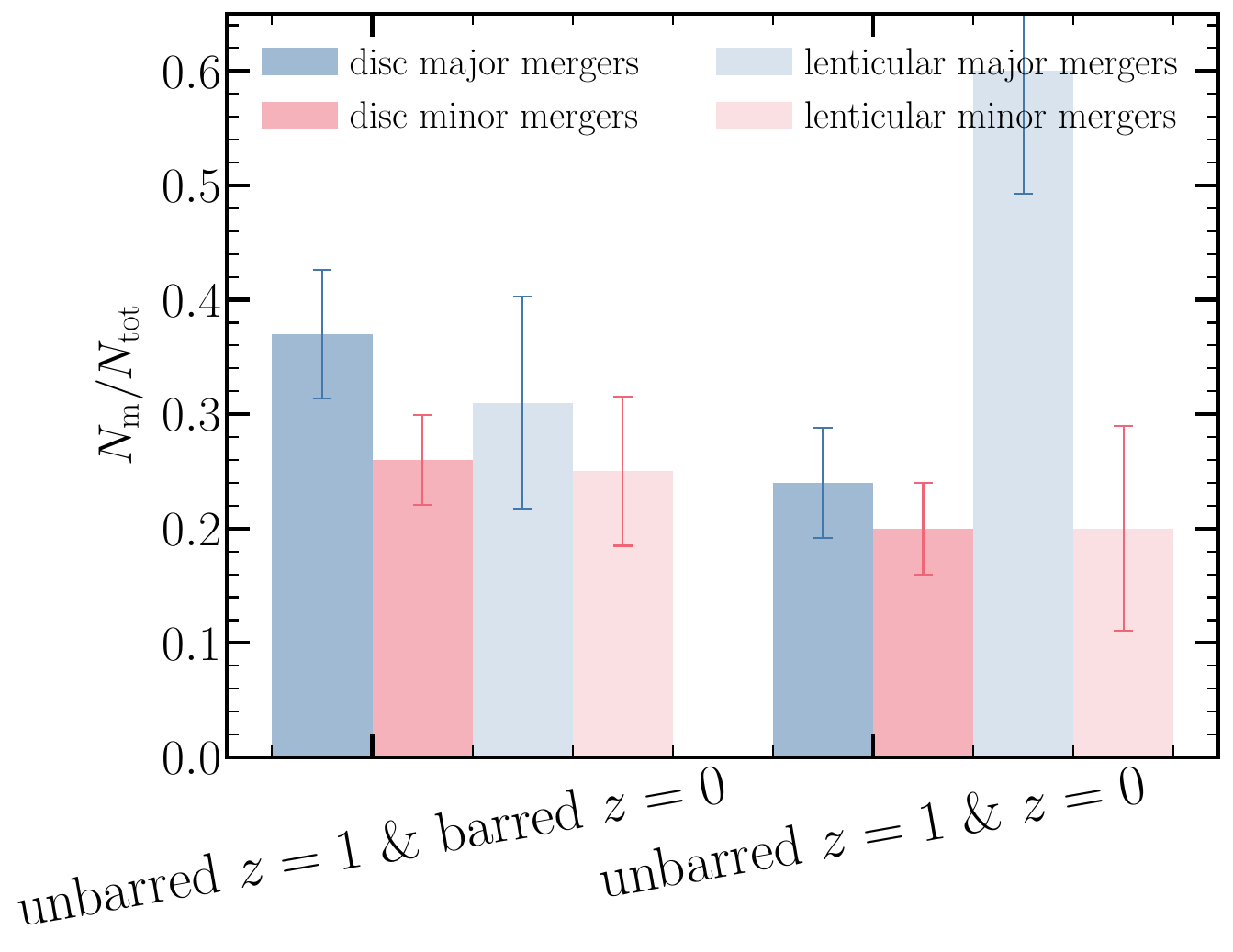}
\caption{Fraction of unbarred galaxies that experienced a major or minor merger since $z=1$ for the disc and lenticular descendants. Error bars represent the Poisson errors of each sample. The highest fraction of major or minor mergers corresponds to those galaxies that form a bar, whereas the lowest ratios are presented by unbarred galaxies that were not able to form a bar.}
\label{fig:mergerunbarred}
\end{figure}


Figure~\ref{fig:mergerunbarred} illustrates the fraction of  disc (darker blue) and lenticular (fainter blue) galaxies that have experienced major or minor mergers since $z=1,$ split into galaxies that form a bar and galaxies that do not form a bar. The figure shows that the lowest merger fractions (less than $0.25$) for disc descendant galaxies are shown by those that have never formed a stable bar. The maximum fraction of galaxies experiencing major mergers is observed when disc galaxies form a bar, with a value of $0.37$.  The fraction of galaxies that have experienced a minor merger exhibit a similar but more moderate trend across all subsamples. In particular, galaxies that are descendants of discs that never formed a bar demonstrate the lowest fractions. Conversely, disc galaxies that develop a bar exhibit the highest fractions.  It should be noted that the fraction of galaxies that experience a minor merger is relatively small. Our findings indicate a possible correlation between the merger history of a galaxy and  the likelihood it will form a bar. Indeed, previous studies proposed that mergers can be another potential bar-formation channel \citep[e.g.][]{lokas2016,martinez2016,lokas2018}.

As demonstrated in previous sections,  lenticular galaxies have more active merger histories than disc galaxies (see Fig.\ref{fig:mergerhistory}), characterised by an elevated fraction of galaxies undergoing major and minor mergers.
However, we observe a similar trend to that typically observed in disc galaxies, except in lenticular galaxies that never developed a stable bar. In this particular case, galaxies that have not developed bars have the highest merger fraction ($0.60$).  This increase may account for the disruption of the disc (as depicted in Fig. \ref{fig:evolutionlentiunbarred}) during later stages of unbarred lenticular descendant galaxies that never developed a bar structure.  

Now, we closely examine galaxies to decipher whether or not these major mergers could trigger bar formation. To this end, we calculated the temporal zone of the merger influence, as stated in Section \ref{subsec:env}. The quantification of this influence can be achieved through the parameter $\eta_{\rm dyn,bf}$, as defined in  Eq. \ref{eq:etadyn}. With this parameter, the discrepancy between the time of bar formation and the time of the most recent major or minor merger or the upcoming major or minor merger is computed relative to the dynamical time of the halo at the moment of bar formation. 

\begin{figure}

\includegraphics[width=1\columnwidth]{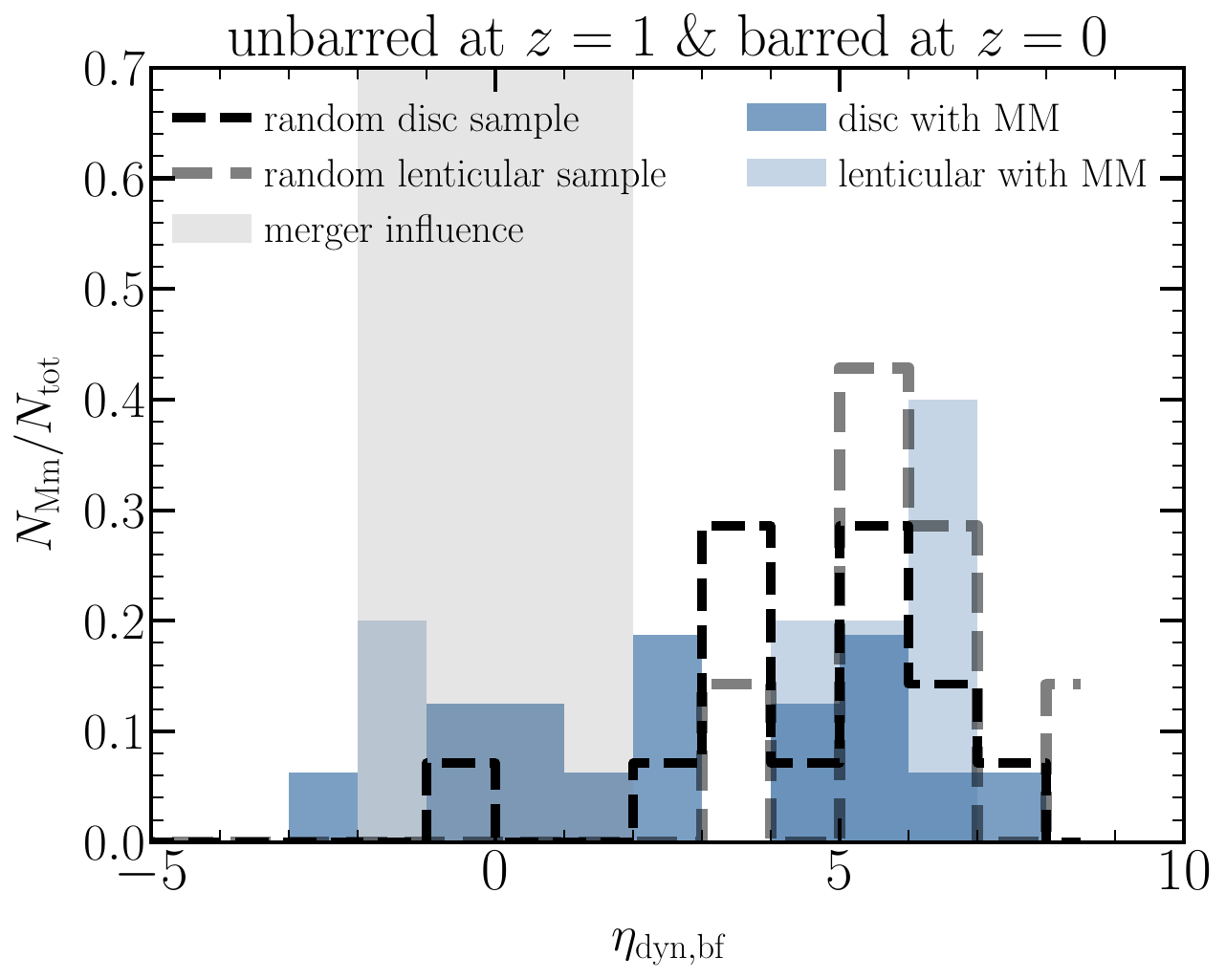}

\caption{Distributions of  $\eta_{\rm dyn,bf}$ defined in Eq. \ref{eq:etadyn} for unbarred galaxies that formed their bar and underwent a major merger. This expression assesses whether
or not the disc galaxy experienced a recent merger at the moment of bar dissolution, and values between $-2$ and $2$ indicate that the galaxy is still influenced by a recent merger. We find that 30 (20) per cent of the disc (lenticular) unbarred galaxies that formed a bar experienced a major merger within less than  two dynamical times.}
\label{fig:tdynuu}
\end{figure}

Figure \ref{fig:tdynuu} depicts the distribution of $\eta_{\rm dyn,bf}$ for descendant galaxies that underwent delayed bar formation and were subjected to at least one significant merger event. Based on the depicted data, it is evident that the minority ---that is, less than $30$ per cent--- of unbarred galaxies that undergo bar formation are influenced by a massive merger event. Similarly, approximately $20$ per cent of lenticular galaxies are affected by a major merger during the period of bar formation. It should be noted that in the context of lenticular galaxies, the occurrence of a recent merger took place slightly after the formation of a bar, as shown by the negative values of $\eta_{\rm dyn,bf}$. This suggests that a small fraction of bars were caused by a merger. In order to establish a causal relationship, we incorporate a random sample of disc galaxies that have undergone a massive merger, but $\eta_{\rm dyn,bf}$ calculated with an identical distribution of the bar formation times. Based on the plot shown in Fig. 15, it can be observed that the random distribution exhibits a peak centred at values of higher than two, which reinforces our suggestion that a connection between bar formation and merger events is only seen in  a small fraction of galaxies. We also explored the distribution of $\eta_{\rm dyn,bf}$ for minor mergers and repeated the same exercise as before, where a similar distinction was observed between the random samples and the descendent galaxies that underwent bar formation when minor mergers were taken into account.

\begin{figure}
\includegraphics[width=1\columnwidth]{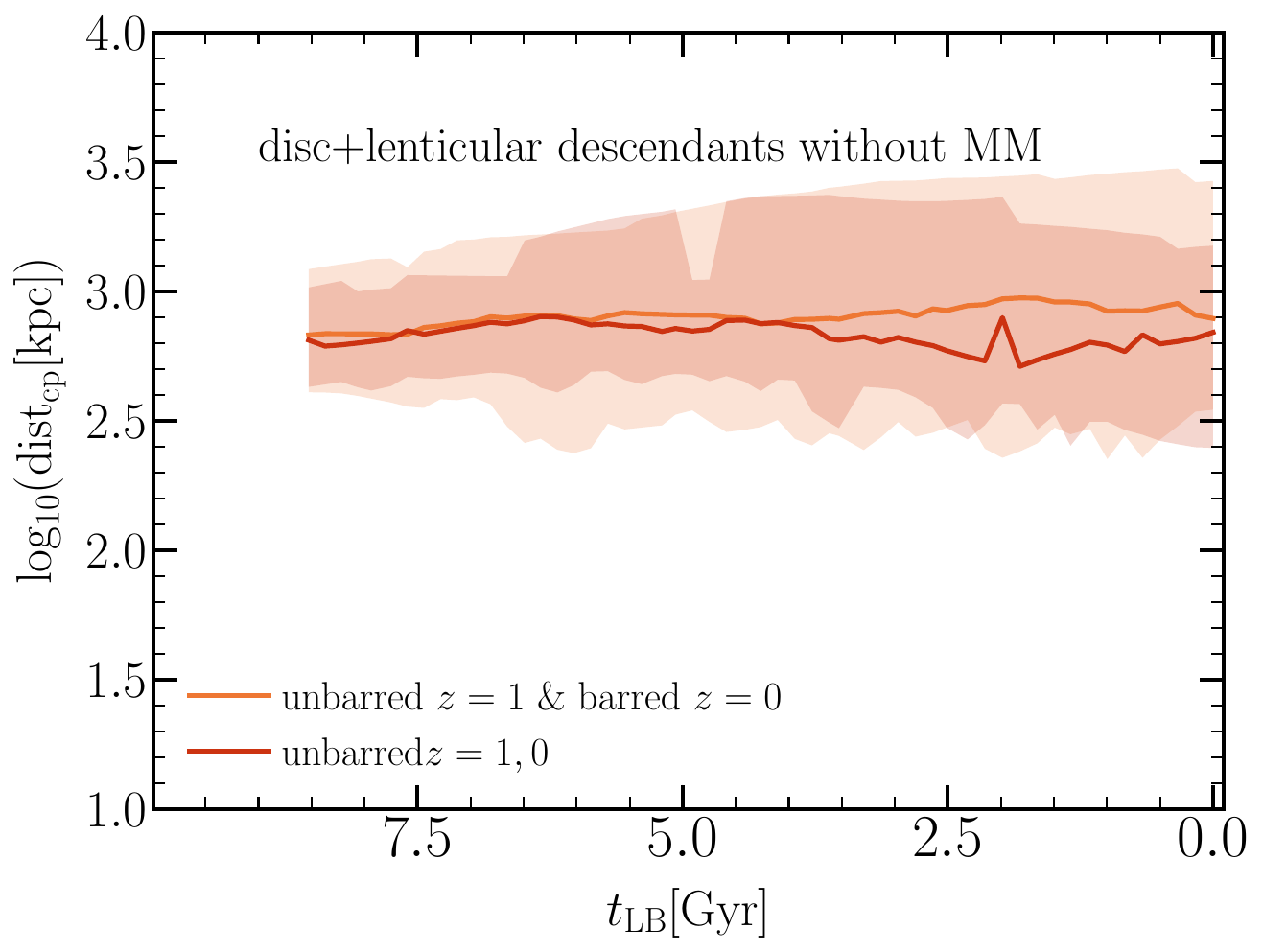}
\caption{Distance to a massive companion (mass ratio $= 1/4$) of disc and lenticular descendant unbarred galaxies that formed bar and did not undergo a major merger (orange line) versus unbarred galaxies that did not undergo a major merger as a function of time since $z=1$. The environment does not seem to not play a role in the likelihood that a disc or lenticular galaxy will form a bar.}
\label{fig:closeneighunbarred}
\end{figure}

We also investigated whether or not tidal interactions with massive neighbours can contribute to the formation of bars.  Figure \ref{fig:closeneighunbarred}  shows the distance to the massive neighbours of these galaxies, and we find that there is no significant difference between unbarred galaxies that develop a bar and those that do not. Also, the satellite fraction as a function of lookback time increases similarly for both categories.

\subsection{Why do some unbarred galaxies not develop a bar?}

\begin{figure*}
\begin{tabular}{c}
\includegraphics[width=2\columnwidth]{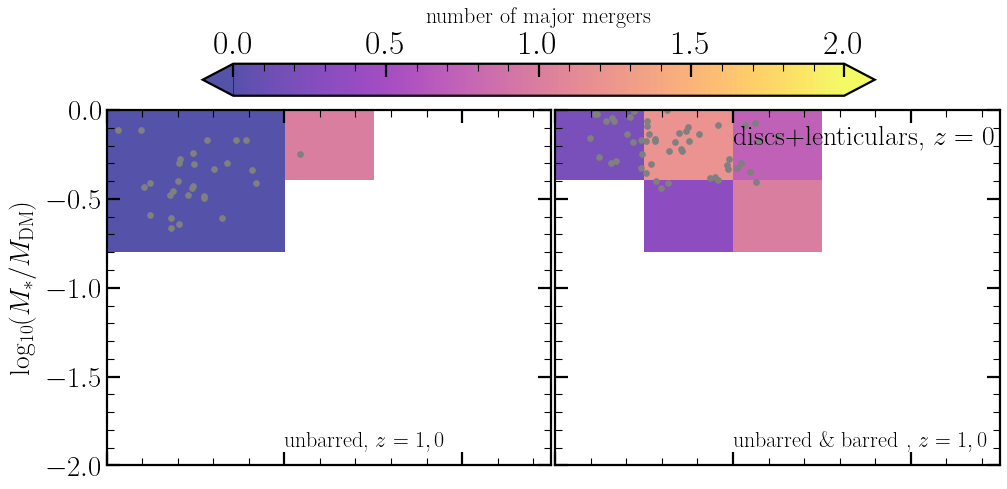} \\
\includegraphics[width=2\columnwidth]{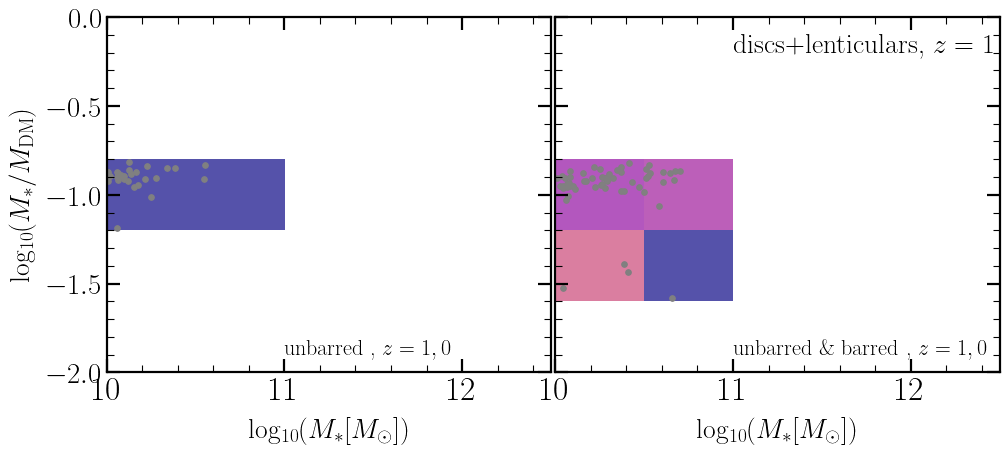} \\
\end{tabular}
\caption{$M_{*}/M_{\rm DM}$ as a function of stellar mass for central disc and lenticular galaxies at $z=1$ and $z=0$. The colour scheme corresponds to the mean number of major mergers for each bin since $z=2$. The top left panel represents the disc descendants that do not develop a bar, whereas the top right panel represents those that develop a bar. The bottom row corresponds to those galaxies at redshift $z=1$. Lenticular and disc galaxies that form a bar after $z=1$ experience more mergers and have higher stellar-to-dark matter ratios than those galaxies that never form a bar for a given stellar mass.}
\label{fig:mstarmhalomstardisc}
\end{figure*}

This section focuses on analysing the galaxy properties associated with the formation of a bar. Specifically, we investigated the stellar mass, gas mass fraction, and the ratio of stellar mass to dark matter mass within a specific aperture (defined as two times the radius of the half stellar mass) for both disc and lenticular descendants. At the present time ($z=0$), galaxies with a stellar bar exhibit characteristics that are  distinct from those of galaxies without bars. These differences include higher stellar mass \citep[e.g.][]{sheth2008}, lower gas fraction \citep[e.g.][]{masters2012}, and dominance of baryonic matter \citep[e.g.][]{fragkoudi2021,rosasguevara2022,izquierdo2022}. The latter disparity is the most significant. We investigate this in more detail in Fig. \ref{fig:mstarmhalomstardisc}.

The diagram illustrates the relation between stellar mass and dark matter in unbarred discs at $z=1$ (shown in the bottom panels), as well as their descendants in the form of discs and lenticular galaxies at $z=0$ (top panels). The galaxies are divided into two groups: those that formed a stable bar (shown in the right column) and those that did not. According to the figure, unbarred galaxies that eventually form a bar tend to have a higher stellar mass than those that do not.  Furthermore, in a general sense, unbarred galaxies have a lower ratio of stellar mass to dark matter compared to galaxies with a bar at $z=0$. This disparity is particularly pronounced among the less massive galaxies ($M_{*}<10^{10.5}\Msun$) in the sample. At $z=1$, the ratio drops for both subsamples. However, there is no substantial difference between galaxies that produce a bar and those that do not. These results are consistent with the findings of \cite*{izquierdo2022}, who conducted a study on a sample of disc galaxies at $z=0$ in the TNG100 and TNG50 simulations, investigating the stellar mass--dark matter mass ratio.   The study consistently found differences between galaxies with and without a bar.  The authors also discovered differences in the stellar mass--dark matter ratio at different radii, even before bars formed. The aforementioned observation is also consistent with the recent findings reported by  \cite*{fragkoudi2021} in their investigation of barred galaxies using the Auriga simulations; these authors established that the stellar component has been the main contributor to the overall rotation curve since $z=0.5$.

Interestingly, Fig. \ref{fig:mstarmhalomstardisc} also shows that galaxies with a bar are more likely to have had an active merger history than galaxies without a bar, thereby supporting our earlier findings (see Fig. \ref{fig:mergerunbarred}). The colour scheme in Fig. \ref{fig:mstarmhalomstardisc} corresponds to the average count of major mergers. Furthermore, at $z=1$, Fig. 17 shows that galaxies that will undergo bar formation after $z=1$ exhibit a higher count of major mergers at early epochs. This result supports the idea that barred galaxies have an earlier assembly in comparison with unbarred galaxies, as shown in  \cite*{rosasguevara2022,rosasguevara2020}. Therefore, the variation in the assembly history of the barred galaxies also affects the structural parameters of their disc and bulge, which in turn influence the creation of a bar. This topic will be investigated in a future project, as it is not relevant to the present discussion.

There is a noteworthy correlation between the presence of a bar and the stellar mass, as shown in Fig. \ref{fig:mstarmhalomstardisc}. This pattern has been identified in multiple studies, most notably in \cite{sheth2008}  using a large sample of discs from the Cosmic Evolution Survey (COSMOS). Nevertheless, the authors found that galaxies with a stellar mass of $10^{10}\Msun$ at $z=1$ have a bar fraction that is comparatively lower than that in the most massive galaxies at this redshift. Conversely, galaxies with a stellar mass of $10^{10}\Msun$ at $z=0$ exhibit a bar fraction comparable to that of the most massive galaxies at $z=0$. This may be in opposition to the results presented in Fig. \ref{fig:mstarmhalomstardisc}, which depicts unbarred galaxies that not only undergo an increase in mass but also fail to form a bar, suggesting that the massive disc ability to effectively form a bar is diminished. As we show in previous sections, this could be attributed to the excess of dark matter in the centre of such galaxies. However, this could also be attributed to the simulation galaxy formation model, which might be deficient in including crucial physical processes at small scales that trigger bar formation. Also, it is important to acknowledge that we are not make a perfect comparison in this study, as we have to take into account the potential bias that the samples may possess. However, it is worth studying this topic further in more detail in future work.


\section{Implications of the environment  on the evolution of the bar fraction}
\label{sec:barfraction}

The bar fraction ---defined as the ratio of the number of disc galaxies with a bar to the total number of disc galaxies at a specific redshift--- and its evolution has the potential to provide insights into the formation and evolution of bars and the assembly of the host galaxies. Several works have looked into the bar fraction, either in observations (e.g.  \citealt{sheth2008,gavazzi2015,melvin2014,cervantes2017, erwin2019}) or simulations (e.g. \citealt{peschken2019}, \citealt{fragkoudi2020}, \citealt{{reddish2022}}). In \cite{rosasguevara2022}, we look into the evolution of the bar fraction in the TNG50. There, we find a gradual evolution in the bar fraction with redshift, increasing from $0.28$  at  $z=4$ and reaching the highest value of $0.48$ at $z=1$.  Subsequently, the bar fraction exhibits a smooth decline to $0.30$ at $z=0$. 

\begin{figure*}
\includegraphics[width=2\columnwidth]{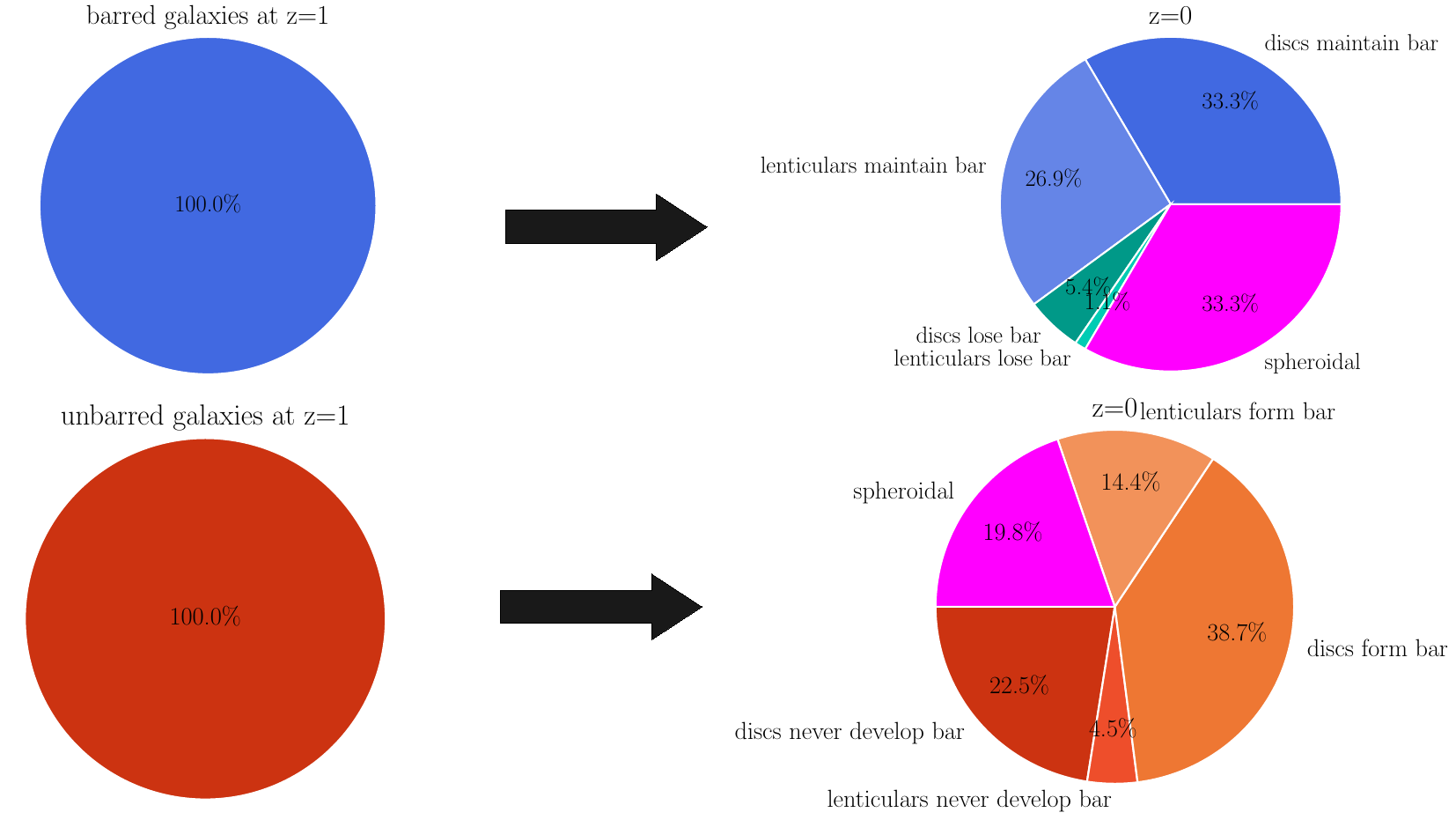}
\caption{Pie chart summarising the fate of barred and unbarred galaxies from $z=1$ to $z=0$. More than 50 per cent of disc galaxies suffer a morphological transformation, whereas just a small percentage
 of galaxies with a stable bar, lose their bar.}
\label{fig:piechart}
\end{figure*}

To understand this evolution, we now look into the evolution of the bar in disc galaxies between $z=1$ and $z=0$. We characterise this evolution in Fig. \ref{fig:piechart}, where we find:
 \begin{itemize}
     \item $33.3$ per cent of the descendants of barred disc galaxies at $z=1$ are discs holding a bar.
     \item $5.4$ per cent are discs losing their bar at $z=0$.
     \item The morphology of the remaining galaxies changes to that of lenticular galaxies, with $26.9$ per cent of them holding a bar and $1.1$ per cent not holding one.
     \item $33.3$ per cent of the barred disc galaxies eventually transform into spheroid galaxies.
     \item  $38.7$ per cent of unbarred galaxies develop a bar at a later time.
     \item  $22.5$ per cent of unbarred galaxies never develop a bar.
     \item $19.8$ per cent and $18.9$ per cent transform into lenticular and spheroid galaxies, respectively.
     \item $4.5$ per cent of lenticular galaxies fail to develop; and the remaining $14.4$ per cent maintain a bar at $z=0$.
 \end{itemize}


Taking into account our findings, we can define the bar fraction, $f_{{\rm bar},zi}$ , at a given redshift $z={i}$ as
\begin{equation}
    f_{{\rm bar},zi} \equiv \frac{n_{{\rm bar},zi}}{n_{{\rm disc},zi}},
\end{equation}
where $n_{{\rm bar},zi}$ and $n_{{\rm disc},zi}$ are the number of disc galaxies with a bar and the number of discs at any given redshift, respectively. In the particular case of the bar fraction at $z=0$, it can be computed as follows:
\begin{eqnarray}
    f_{{\rm bar},z0} &=& \frac{n_{{\rm bar},z0}}{n_{{\rm disc},z0}} \\
                     &=& f_{{\rm bar},z1} + \sum_{z=1}^{z=0} \frac{\Delta f_{{\rm bar},z1}}{\Delta t } \Delta t
\label{eq:fbar}
\end{eqnarray}
The second term, which is the change of the bar fraction between $z=1$ and $z=0$, is contingent on the formation and dissolution of bars during this period of time. This term can be subdivided into six distinct components, including the morphological transformation of galaxies, as outlined below:
\begin{eqnarray}
 \sum_{z=1}^{z=0} \frac{\Delta f_{{\rm bar},z1}}{\Delta t } \Delta t &=&  f_{{\rm bar form},z<1}+   f_{{\rm bar form, disc form} z<1} \notag \\
    &-& f_{{\rm bar lost},z<1}  - f_{{\rm bar,des discs},z<1} \notag \\
    &-& f_{{\rm bar lost, disc,formed}, z<1} \notag \\
    &-&f_{{\rm bar, disc, form, des}, z<1}
    \label{eq:changebarfraction}
\end{eqnarray}
The initial two terms exhibit positive values and are indicative of the emergence of bars after $z=1$. Conversely, the subsequent terms have negative values, which indicate the dissolution of a bar or the destruction of a disc.

The first term of Eq. \ref{eq:changebarfraction} corresponds to the fraction of bars found on one of our disc descendant subsamples. More precisely, the subsample corresponds to the descendant galaxies that have a disc with a dominant cold component at $z=1$, but with a bar forming after $z=1$. The given fraction can be represented as $f_{{\rm bar form},z<1}= (n_{{\rm bar,form}}/ n_{{\rm disc},z1}) (n_{{\rm disc},z1}/n_{{\rm disc},z0}),$ where $n_{{\rm disc},z0}=349$ represents the number of discs at $z=0$  in the catalogue of \citet{rosasguevara2022}; $n_{{\rm disc},z1}=204$  represents the number of galaxies with a cold component, which corresponds to the number of galaxies in our parent sample (see Table \ref{table:morphology}) and $n_{{\rm bar,form}}=43$ the number of bars that formed after $z<1$ of this subsample of descendant discs (see Table \ref{table:morphology}). The value of  $f_{{\rm bar form},z<1}$ can be calculated as  $43/204\times  204/349= 0.1$.   It is worth noting that a fraction of this value, namely approximately $30$ per cent, is attributed to environmental factors, specifically major or minor mergers. This observation is supported by the findings presented in subsection \ref{subsec:unbarredmergers} and illustrated in Fig. \ref{fig:mergerunbarred}  and Fig. \ref{fig:tdynuu}. These figures indicate that the contribution of environmental factors (mergers) to the formation of new bars in the bar fraction ($f_{{\rm bar form},z<1}$) is limited to a maximum of $6$ per cent.

The second term, $f_{{\rm bar form, disc form} z<1}$, corresponds to the bars formed in discs that have finished assembling the large cold-disc component at a redshift of lower than $1$. The expression may be represented as $f_{{\rm bar form, disc form} z<1}= (n_{{\rm bar,form}<z1}/ n_{{\rm disc},z<1}) (n_{{\rm disc},z<1}/n_{{\rm disc},z0}),$ where  $n_{{\rm disc},z<1}=349-204=145$. This is the number of discs that the cold-disc component formed after $z=1$. The number of bars formed in these discs is represented as $n_{{\rm bar form}<z1}=105-74$. The first number represents the number of bars observed in disc galaxies at $z=0$ and comes from the catalogue of \citealt[][see their Table 2]{rosasguevara2022} and the second number is the number of bars of our disc descendant subsamples and is given in Table \ref{table:morphology}. This gives us the value of $f_{{\rm bar form, disc form} z<1}=0.1$.  Assuming that $30$ per cent of the bars formed are a consequence of mergers (the same fraction identified in bars formed in the descendants of discs at $z=1$), it can be inferred that approximately $6$ per cent of the bars in the discs may be attributed to a major or minor merger event. This implies that a maximum of $12$ per cent of the bars at $z=0$ that formed after $z=1$ can be attributed to environmental influences.  It should be noted that the aforementioned value of $12$ per cent serves as an upper bound. This is due to the assumption that the rate of mergers for the discs produced after $z=1$ remains constant. However, it is important to acknowledge that this assumption is  not entirely accurate, as it is well established that the merger rate for a given mass falls over time (see Fig 4 in \citealt{rodriguezgomez2015}).

The final four terms in Eq. \ref{eq:changebarfraction} are associated with the process of bar dissolution and exhibit negative values.
The third term, $f_{{\rm bar lost},z<1}$, represents the fraction of bars that have dissolved in barred disc galaxies at $z=1$. This corresponds to our subsample of disc galaxies that were barred at $z=1$ and afterwards experienced the dissolution of that bar. The calculation involves the product of two ratios $f_{{\rm bar lost},z<1}= (n_{{\rm bar, lost}}/n_{{\rm disc},z1})(n_{{\rm disc},z1}/n_{{\rm disc},z0})= 5/204 \times 204/349= 0.01$.  The first ratio, $ (n_{\rm bar, lost}/n_{{\rm disc},z1})$, represents the number of galaxies with dissolved bars divided by the number of disc galaxies at $z=1$.  The second ratio is the number of discs at $z=1$ over the number of discs at $z=0$. As we have seen in section \ref{sec:environ} and Figs. \ref{fig:tdynbu} and \ref{fig:closeneigh}, these galaxies suffer a tidal influence from a massive companion in temporal proximity to the dissolution of the bar. Additionally, it is noted that these galaxies tend to inhabit environments characterised by higher density. Considering the available evidence, it may be inferred that the aforementioned $1$ per cent can be attributed to environmental influences.

The fourth term of Eq.~\ref{eq:changebarfraction}, $f_{{\rm bar, des disc},z<1}$, represents the fraction of bars in the descendants of disc galaxies at $z=1$ that experience damage or destruction of their cold-disc component, resulting in a morphological transition. We can express $f_{{\rm bar,des disc},z<1}=( n_{{\rm bars},z1}/n_{{\rm nondisc},z0})(n_{{\rm non disc},z0}/ n_{{\rm disc},z1})(n_{{\rm disc},z1}/n_{{\rm disc},z0})$. This fraction can be calculated by finding the number of barred galaxies in our sample that transform into lenticular and spheroidal galaxies, which corresponds to $26$ and $31,$ respectively (see Table \ref{table:morphology}) and the number of the total  galaxies that transform into lenticulars and spheroidals ($47$ and $53$). The combined values in $f_{{\rm bar, des discs},z<1}$ are determined to be $0.16$.

The fifth term, denoted $f_{{\rm bar lost, disc,form}, z<1}$, is aligned with the third term and represents the fraction of dissolved bars in discs where the formation of their cold-disc component occurs after $z=1$.
The given expression may be represented as $f_{{\rm bar lost, disc,form}, z<1}= (n_{{\rm bar,lost,disc form} z<1}/ n_{{\rm disc,form}, z<1})$ $(n_{{\rm disc,form},z<1}/n_{{\rm disc},z0})$.  The analysis conducted here does not incorporate the dissolution of the bars in the discs that formed a dominant cold component beyond $z=1$. However, it is reasonable to assume that similar processes, such as environmental factors, may contribute to this dissolution with comparable efficiency. Consequently, the ratio $(n_{{\rm bar,lost,disc,form} z<1}/ n_{{\rm disc,form}, z<1})$  is equal to $0.01$ and $(n_{{\rm disc,form},z<1}/n_{{\rm disc},z0})= 145/349$.  By substituting the appropriate values, we determine that $f_{{\rm bar lost, disc,form}, z<1}=0.004$.

Lastly, the final term, $f_{{\rm disc, form, des}, z<1}$, represents the fraction of discs that underwent the formation of their cold component after $z=1$, forming a bar and then experiencing the destruction of the disc. We assume that this fraction is comparable to that of the descendant galaxies whose cold-disc component undergoes damage or complete destruction, leading to a morphological transition. In such cases, the calculated value for $f_{{\rm bar, disc, form, des}, z<1}$ will be aligned with the fourth term of Eq. \ref{eq:changebarfraction}, which is $0.16$. 


Gathering our results, we estimate the product of Eq. \ref{eq:changebarfraction}  to be as follows:
\begin{eqnarray*}
 \sum_{z=1}^{z=0} \frac{\Delta f_{{\rm bar},z1}}{\Delta t } \Delta t &=&  0.1+0.1-0.01-0.16-0.004-0.16 \\
 &=& -0.13.
 \end{eqnarray*}
The bar fraction of our parent sample at $z=1$  is $0.45\pm 0.03$ and based on this analysis and Eq. \ref{eq:fbar}, the bar fraction that we obtain at $z=0$ is $0.31$, whereas the bar fraction at $z=0$ is $0.30\pm0.02$, which is in good agreement, and the possible discrepancy that we have could be linked to the assumptions for the discs that formed later than $z=1$. We also note that this gave us information about the galaxies that assembled their massive cold-disc component after $z=1,$ because the number of discs with a dominant cold component increases with decreasing redshift (see Table 1 in \citealt{rosasguevara2022} and \citealt{pillepich2019,vanderwel2014}). This suggests that the net efficiency of forming bars could be lower in discs that assembled later in time, because there is a slight decrease in the bar fraction.


\subsection{Outlook}

The simulation TNG50 allows us to estimate the importance of the environment in the evolution of the bar fraction. However, we must acknowledge the constraints associated with this estimation. For instance, we focus on the bar fraction from $z=1$ to $z=0$ as well as for massive disc galaxies ($M_{*}>10^{10}\Msun$). Nevertheless, we find bars in massive galaxies since $z=4$ in the simulation \citep{rosasguevara2022}, which aligns with recent observations from the JWST at such high redshifts \citep{guo2023,tsukui2023,leconte2023}. The conditions experienced by galaxies at such high redshifts are notably distinct from those at lower redshifts, where it is anticipated that the frequency of encounters and mergers is higher. In light of this, it is imperative to investigate the influence of the surrounding environment on the bar fraction at lower stellar masses and higher redshifts, as has been suggested by \cite{costantin2023}, who, for example, found a bar in a low-mass disc galaxy ($3.9 \times 10^9\Msun$ at $z=3$). Indeed, \cite{zana2022} find a small bar fraction (less than 0.1, see their Fig. 17) in TNG50 disc galaxies with stellar masses of $10^9$--$5\times10^9 \Msun$.   However, the authors do not conduct an exhaustive investigation to determine whether these bars represent transient structures or exhibit temporal stability.  Alternatively, further investigations into low-mass disc galaxies ---as reported in \cite{costantin2023}--- could give us new constraints of the bar fraction in disc galaxies in the low-mass range.

In addition, the present analysis does not consider the potential weakening or disappearance of bars due to vertical instabilities in discs formed after $z=1$. However, earlier research using N-body simulations demonstrated the ability of bars to reform under such circumstances \citep{bournaud2005,berentzen2004}.

An additional factor that we do not see in discs formed at $z=1$ but that could affect a disc formed after $z=1$ is  the dissolution of bars due to the tidal interactions resulting from minor mergers. An illustrative instance can be observed in the study conducted by \cite{ghosh2021}, wherein N-body simulations were employed to investigate the impact of tidal interactions resulting from minor mergers on the central stellar bar in the remnant of the merger. Following each pericentre transit of the satellite, the central bar undergoes temporary phases of bar amplification. However, the primary occurrence of bar weakening occurs only after the merger event. Nevertheless, it is plausible that a galaxy could undergo the process of accreting cold gas either during mergers or at a subsequent stage, thus potentially revitalising the bar structure.   \cite{ghosh2021} showed that minor mergers do not provide any discernible influence. However, it is worth noting that this phenomenon may hold significance in scenarios characterised by lower redshift or high-density conditions, where galaxies are subject to more exposed tidal interactions.

Tidal interactions may play a significant role in the creation and dissolution of bars in galaxies with lower stellar masses and in highly dense environments. For example, the studies conducted by \cite{lokas2014} and  \cite{lokas2016} examine the process of bar evolution in dwarf galaxies and Milky Way galaxies resulting from tidal interactions within a cluster-like setting. The authors discovered that the characteristics of the bars exhibit temporal variations and are subject to the influence of the magnitude of the tidal force encountered during the evolutionary process. The formation of bars in galaxies is observed to occur at earlier stages and the bars themselves are longer and stronger in galaxies that experience a higher degree of tidal force exerted by the surrounding cluster.  This particular form of interaction has not been extensively examined; however, it is our intention to explore this topic in a forthcoming research project.

AGN feedback has been identified as an additional mechanism linked to the formation and dissolution of a bar. In particular, \cite{bonoli2016} find that a simulated galaxy forms a strong bar below $z\sim1$, and the authors point out that the disc in the simulation is more prone to instabilities compared to the original \textit{Eris}, possibly because of early AGN feedback affecting the central part of the galaxy. \cite{zana2018c}, studying an enhanced suite of \textit{Eris}, highlight the effects of the feedback processes on the formation time and final properties of the bar. In a study conducted by \cite{lokas2022}, it was observed that a significant merger event resulting in the coalescence of two supermassive black holes led to an enhancement of the AGN feedback within the TNG 100 simulation. This, in turn, caused the expulsion of gas from the central region of the galaxy, subsequently suppressing star formation and facilitating the formation of a bar structure. However, it is worth noting that AGN feedback may potentially exert a counteractive influence by weakening the bar and potentially resulting in their eventual removal \citep{irodotou2022}. This process is a potential channel to understand the formation and dissolution of a bar. We identified a specific instance in which a bar disappeared following a significant interaction with a companion, with a subsequent, notable episode
of strong AGN feedback. Nevertheless, it remains uncertain as to whether the tidal interaction is responsible for the disappearance of the bar and the subsequent activation of the AGN, or if the AGN feedback itself plays a role in the dissolution of the bar.

One crucial consideration to be mindful of is the constrained volume of TNG50, which presents limitations in terms of its ability to allow comprehensive investigations of a wide range of diverse environments, particularly for disc galaxies in a wider range of stellar masses. This is another aspect that may see considerable development  in the future with the next generation of cosmological hydrodynamical simulations.


\section{Summary}
\label{sec:summary}

In this study, we analysed the impact of the environment on the evolution of barred and unbarred disc galaxies at $z=1$.  We used the TNG50 simulation, which is a magneto-cosmological hydrodynamical simulation based on the $\Lambda$CDM model as part of the IllustrisTNG project \citep{pillepich2019,nelson2019b}.

As part of the approach taken here, the galaxy components, such as the classical bulge, thin disc, and thick disc, are identified using the kinematic decomposition software \textsc{mordor} \citep{zana2022} and the identification of the bar structures is done using the Fourier decomposition of the face-on stellar surface density.

To quantify the environment, we investigated the merger histories of disc galaxy descendants and identified the temporal region of influence of major mergers in the galaxy. We also explored the distance to the closest massive companion to identify the possible tidal interactions that  galaxies experience, and, finally, we investigated the number of satellites as a function of redshift to assess the large-scale environment.

Our findings can be summarised as follows:
\begin{itemize}
\item A significant fraction of disc galaxies with a stellar mass of $M_{*}\gsim 10^{10} \Msun$ undergo a morphological change between  redshift $z=1$ and $z=0$. Specifically, $49$ per cent of these galaxies evolve into lenticular or spheroidal galaxies, as indicated in Table \ref{table:morphology}, while the remaining $51$ per cent retain their disc morphology.

\item Those galaxies that transform into lenticular and spheroidal galaxies have an active (major) merger history (see Fig. \ref{fig:mergerhistory} and Table \ref{tab:discmergers}). The descendant spheroidal and lenticular galaxies exhibit the highest fraction of galaxies that experience at least one major merger ($0.36$ and $0.38$), while the descendant disc galaxies have the lowest fraction ($0.15$) since $z=1$. Furthermore, spheroidal galaxies generally exist in a denser environment compared to other galaxies, as they possess the highest fraction of satellite galaxies over time.
\end{itemize}

To explore the rise and decline of bar structures, we divided the parent disc sample into two categories: barred disc galaxies at $z=1$ and unbarred disc galaxies at the same redshift. Subsequently, we monitored the development of these two groups. Concerning the barred disc galaxies at $z=1$, our findings indicate that:

\begin{itemize}
 \item  $5.4$ and $1.1$ per cent of disc and lenticular descendants, respectively, lose their bar (see  Table \ref{table:morphology} and  Fig. \ref{fig:piechart}). On the other hand, $33.3$ and $26.9$ per cent of the barred disc and lenticular descendants, respectively, maintain their bars. The remaining $33.3$ per cent become spheroidal galaxies.

\item There are variations in the characteristics of the bars: the galaxies that are capable of retaining their bars exhibit the strongest and most elongated bars, whereas the galaxies whose bar dissolves overtime overall tend to have weaker and shorter bars to start with at $z=1$ (see Figs. \ref{fig:evolutiondiscs} and \ref{fig:evolutionlenti}). The subsample that preserves its bar structure has its stellar disc dominated by the thin-disc component.



\item Barred disc and lenticular galaxies that have lost their bar have a more active history of mergers than those that have retained their bar (Fig. \ref{fig:mergersbar}). We observe that in galaxies experiencing a major merger (0.33) and bar loss after $z<1$, the merger takes place near the dissolution time. However, this does not hold for lenticular galaxies.

\item For galaxies that lose their bar but did not undergo a major merger, we find that they have a close encounter with a massive companion compared to galaxies that maintain their bar (Fig. \ref{fig:closeneigh}). Furthermore, the satellite fraction of disc and lenticular galaxies losing their bar increases from $0.15$ at $z=1$ to $0.5$ at $z=0$, which is significantly lower than the increase from $0.0$ to $0.3$ observed in galaxies that retain their bar (see Fig. \ref{fig:fracsatellitetypebars}).

\end{itemize}

Concerning the evolution of unbarred disc galaxies at $z=1$, we find the following:

\begin{itemize}

\item  $53.1$  per cent of unbarred descendant galaxies develop a bar (Table \ref{table:morphology} and Fig. \ref{fig:piechart}), whereas $27.1$ of them do not develop a stable bar at $z=0$ and $19.8$ per cent transform into a spheroidal galaxy.

\item  In disc galaxies that experience bar formation, the bar structure typically emerges within the first few gigayears after $z=1$ (see Fig. \ref{fig:evolutiondiscsunbarred}). These bars are relatively weak and short compared to the bars of disc galaxies that formed a bar at earlier times. This observation aligns with the notion that the bars in the former galaxies formed at an earlier stage  \citep{rosasguevara2020,izquierdo2022}. The lenticular descendants form the bar before undergoing morphological transformation, and the disc has not yet been destroyed (see Fig. \ref{fig:evolutionlentiunbarred}).


\item Unbarred disc descendants that fail to create a stable bar exhibit the lowest galaxy fraction that has undergone major mergers ($0.24$, see Fig.  \ref{fig:mergerunbarred}). Conversely, unbarred disc descendants that eventually form a stable bar have the highest galaxy fraction  ($0.36$) that has undergone major mergers, excluding those that transform into lenticular galaxies and never undergo bar formation.


\item Galaxies that form a bar after $z=0$ show a temporal correlation with a recent major merger. 
Our results show that at least $30$ per cent and $20$ per cent of the disc and lenticular galaxies, respectively, form their bars during the influence of the major merger (see Fig. \ref{fig:tdynuu}). We find similar results when we look at the time when the minor mergers happened.


\item We find that, on average, there is no difference in the distances to the closest massive companion between galaxies that form a bar and those that do not (see Fig. \ref{fig:closeneighunbarred}).


\item On average, galaxies that form a bar structure exhibit a greater stellar mass \citep[e.g.][]{gavazzi2015,consolandi2016}, a reduced gas fraction \citep[e.g.][]{masters2012,kruk2018}, and an elevated ratio of stellar mass to dark matter mass \citep[e.g.][]{fragkoudi2021,izquierdo2022}. There is a possible connection between the evolution of this latter ratio and the merger history both before and after $z=1$. We find a disparity in the ratio of stellar mass to dark matter mass for a specific stellar mass. This disparity is already higher in barred galaxies, even before the formation of the bars. Additionally, unbarred galaxies that form a bar after $z=1$ undergo a higher number of major mergers compared to unbarred galaxies that never develop a bar. This suggests that the formation of barred galaxies occurs before the formation of galaxies that never generate a bar.
\end{itemize}

In summary, we find that bars are prevalent in disc galaxies and are stable. Nevertheless, bar evolution can be influenced by the surrounding environment. Tidal interactions resulting from mergers or encounters with close companions have the potential to disrupt the central region of the disc. This disruption can manifest as either the dissolution of the bar, the formation of a bar, or, in the most severe case, the complete destruction of the disc, which causes a morphological transformation of the galaxy. Weak and short bars at $z=1$ that enter crowded environments or experience a merger may dissolve according to our findings. Unbarred galaxies at $z=1$ that undergo major or minor mergers may eventually develop a bar. The influence of a merger or close encounter on the formation or dissolution of a bar is dependent on the orbital parameters, which could be investigated in future studies through the combination of cosmological hydrodynamic simulations and high-temporal-resolution simulations. In conclusion, our findings suggest that the environment may have influenced the evolution of the bar fraction, a common observable used to understand the evolution of bars.


\begin{acknowledgements}
The authors thank the referee for the useful comments and the TNG team for early access to the TNG50 data prior to its public release. Y.R.G. acknowledges the support of the``Juan de la Cierva Incorporation'' Fellowship (IJC2019-041131-I). Y.R.G. thanks Karin Menendez-Delmestre and the Conference for Galactic Bars: Driving and Decoding Galaxy Evolution for the useful discussions. S.B. acknowledges support from the Spanish Ministerio de Ciencia e Innovación through project PID2021-124243NB-C21. D.I.V. acknowledges the financial support provided under the European Union’s H2020 ERC Consolidator Grant ``Binary Massive Black Hole Astrophysics'' (B Massive, Grant Agreement: 818691).
\end{acknowledgements}

%
%

\begin{appendix}

\section{Calculation of the dissolution time of a bar}
\label{append:disolution}
\begin{figure}
\begin{tabular}{c}
\includegraphics[width=1\columnwidth]{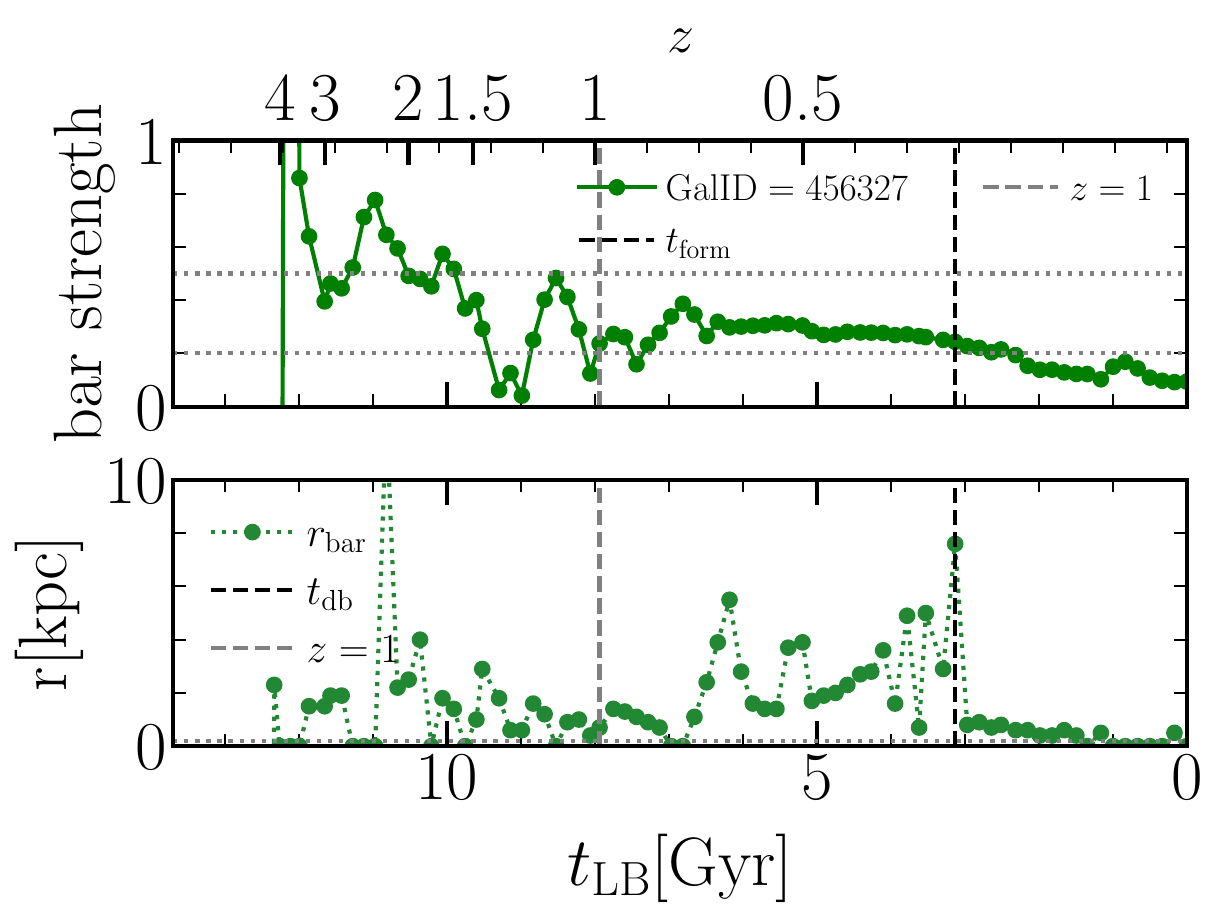}
\\
 \includegraphics[width=1\columnwidth]{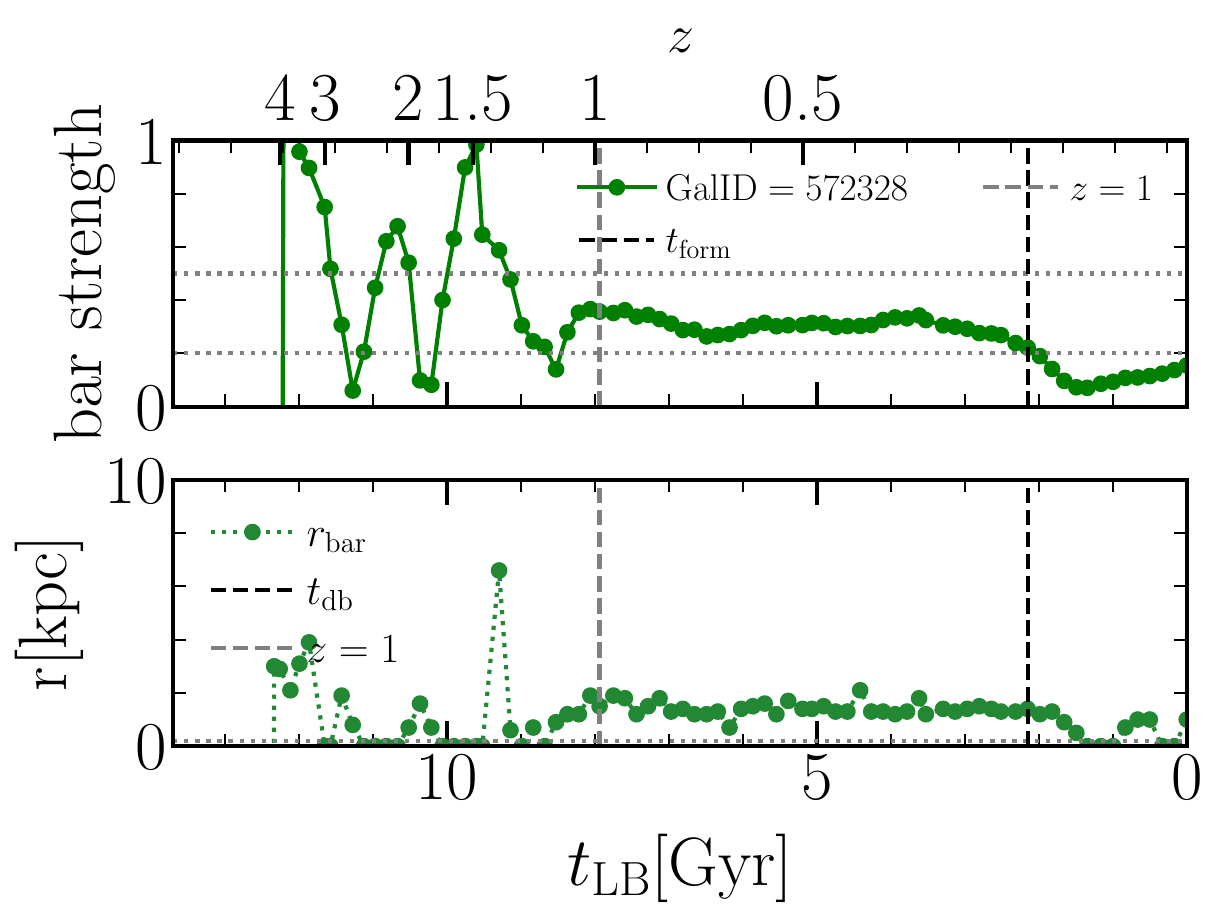}\\
\end{tabular}
\caption{Examples of the bar evolution of disc galaxies at $z=1$ that dissolved their bar. The black solid line indicates the  calculated time of the dissolution of the bar, and the grey solid line represents the lookback time that corresponds to $z=1$.}
\label{fig:barev_dissolution}
\end{figure}

In this section, we present some examples of disc galaxies that dissolved their bar and the time of bar dissolution, as explained in subsection \ref{sub:discsample}. The time dissolution is determined by the lookback time when the bar strength is smaller than 0.2 and the bar length is smaller than the minimum radius imposed due to spatial resolution.  Figure \ref{fig:barev_dissolution} shows two examples of the bar strength and bar length of disc galaxies that were barred at $z=1$ and dissolved their bar thereafter.
\end{appendix}
\end{document}